\documentclass[
 reprint,
 amsmath,
 amssymb,
 aps,
 prx,
]{revtex4-2}

\usepackage{graphicx}
\usepackage{dcolumn}
\usepackage{bm}
\usepackage{booktabs}
\usepackage{makecell}
\usepackage{subfigure}
\usepackage{xcolor}
\usepackage{footmisc}
\usepackage[colorlinks,linkcolor={blue!80!black},citecolor={blue!80!black},urlcolor={blue!80!black}]{hyperref}
\usepackage{amsmath}
\usepackage{multirow}
\usepackage[normalem]{ulem}
\usepackage{braket}

\begin{document}

\preprint{APS/123-QED}

\title{Mitigating Capacitive Loading Enables Fast Two-Qubit Gates in Highly Connected Fluxonium Quantum Processors}

\author{Quan Guan}
\author{Guo Xuan Chan}
\author{Xu Dou}

\author{Chunqing Deng}
 \email{dengchunqing@quantumsc.cn}
 
\author{Lijing Jin}
 \email{jinlijing@quantumsc.cn}

\affiliation{Quantum Science Center of Guangdong-Hong Kong-Macao Greater Bay Area, Shenzhen, China}

\date{\today}

\begin{abstract}
Fluxonium qubits combine long coherence times with strong anharmonicity, making them attractive for scalable superconducting quantum processors. Recent experiments have demonstrated high-fidelity two-qubit gates and multi-qubit entanglement in one-dimensional and connectivity-four fluxonium processors, motivating their extension to highly connected two-dimensional (2D) architectures. A central challenge in this extension is to achieve strong coupling to multiple circuit elements while preserving the finite capacitance budget of fluxonium qubits, which constrains the attainable qubit–qubit effective coupling strength and, consequently, two-qubit gate performance. 
Here, we establish a unified theoretical framework that identifies the connectivity-dependent upper bound on capacitive coupling and quantifies its reduction by distinct parasitic-capacitance channels, applicable across both fluxonium and transmon regimes.
We show that pad-to-ground loading is primarily geometry limited and can be substantially suppressed through qubit-pad engineering, whereas inter-pad loading is ultimately constrained by parasitic capacitances of Josephson junctions and Josephson junction arrays. Building on these insights, we formulate practical design principles and numerically demonstrate two-qubit gates as fast as 27 ns and leakage-induced infidelity below $10^{-3}$ under representative fabrication variations in 2D fluxonium architectures. 
These results establish a quantitative framework for understanding and mitigating capacitive loading, suggesting that it does not impose a fundamental performance limit on highly connected fluxonium processors.
\end{abstract}

\maketitle

\section{Introduction}
Superconducting circuits, particularly transmon-based architectures~\cite{Koch2007_transmon}, have achieved major milestones ranging from demonstrations of quantum computational advantage~\cite{GoogleQAI_2019, Zuchongzhi3.2_2025} to realization of logical error suppression toward fault-tolerant quantum computation~\cite{Krinner2022_IBMETH_d3, GoogleQAI_2023, GoogleQAI_2025}. Beyond transmon, fluxonium qubits have emerged as an alternative for scalable quantum computation owing to their millisecond-scale coherence times, high-fidelity operations, and intrinsic suppression of residual static and spectator interactions within the computational subspace~\cite{Manucharyan2009_fluxonium,Somoroff2023, Fei2025,Ding2023_MIT_FTF}. Recent experiments have further demonstrated parallel operations and multi-qubit entanglement in a one-dimensional (1D) fluxonium processor with tunable-couplers~\cite{Zhan2026_22q_GHZ}, and a connectivity-four fluxonium–resonator–fluxonium unit cell~\cite{Schirk2026}, providing the foundation for scaling fluxonium qubits to highly connected 2D architectures.

At higher qubit connectivity, a fundamental challenge emerges: achieving strong capacitive coupling while maintaining a finite shunt capacitance. Compared with conventional transmon qubits, fluxonium qubits typically operate within a relatively small capacitance budget—a finite electrostatic resource shared among all desired and parasitic capacitive channels. The desired larger capacitive couplings and additional circuit elements progressively consume the available budget, reducing the achievable coupling strength. This phenomenon, referred to as capacitive loading~\cite{Rosenfeld2024, Heunisch2025, Peng2026}, has therefore become an increasingly relevant constraint for highly connected 2D fluxonium quantum processors. This challenge becomes particularly acute for fast microwave-activated phase (MAP) gates~\cite{Ding2023_MIT_FTF}, which rely on exceptionally strong qubit–coupler coupling, and is further exacerbated in floating or double-mode coupler architectures~\cite{Chan2026_F_DTC}, which require larger capacitive couplings to achieve the same interaction strength as in grounded coupler designs~\cite{Ding2023_MIT_FTF}. While previous studies have recognized this constraint and explored architecture-specific mitigation strategies—such as floating configurations and high-impedance couplers~\cite{Ding2023_MIT_FTF,Rosenfeld2024,Peng2026, Schirk2026}—the physical origin and a unified quantitative framework of capacitive loading in superconducting devices remain lacking.

In this work, we develop a quantitative framework for capacitive loading in capacitively coupled superconducting circuits by deriving an analytical relation between the qubit capacitance budget and its achievable coupling to external circuit elements. The resulting capacitance-budget law separates the effects of qubit connectivity, parasitic capacitance, geometric design, and fabrication constrain on the attainable coupling strength, providing a unified basis for understanding and mitigating capacitive loading. In addition to providing a physical interpretation of capacitive loading mitigation in previous studies, we identify the fabrication-induced parasitic capacitances of Josephson junctions and Josephson junction arrays as an intrinsic limitation, while demonstrating that qubit-pad geometry provides an effective degree of freedom for mitigating geometry-dependent loading. These insights establish practical design principles for scaling capacitive coupling with connectivity and enable fast two-qubit gates with leakage-induced infidelity below $10^{-3}$ under representative fabrication variations.

\begin{figure*}[!t]
    \centering \includegraphics[width=0.95\linewidth]{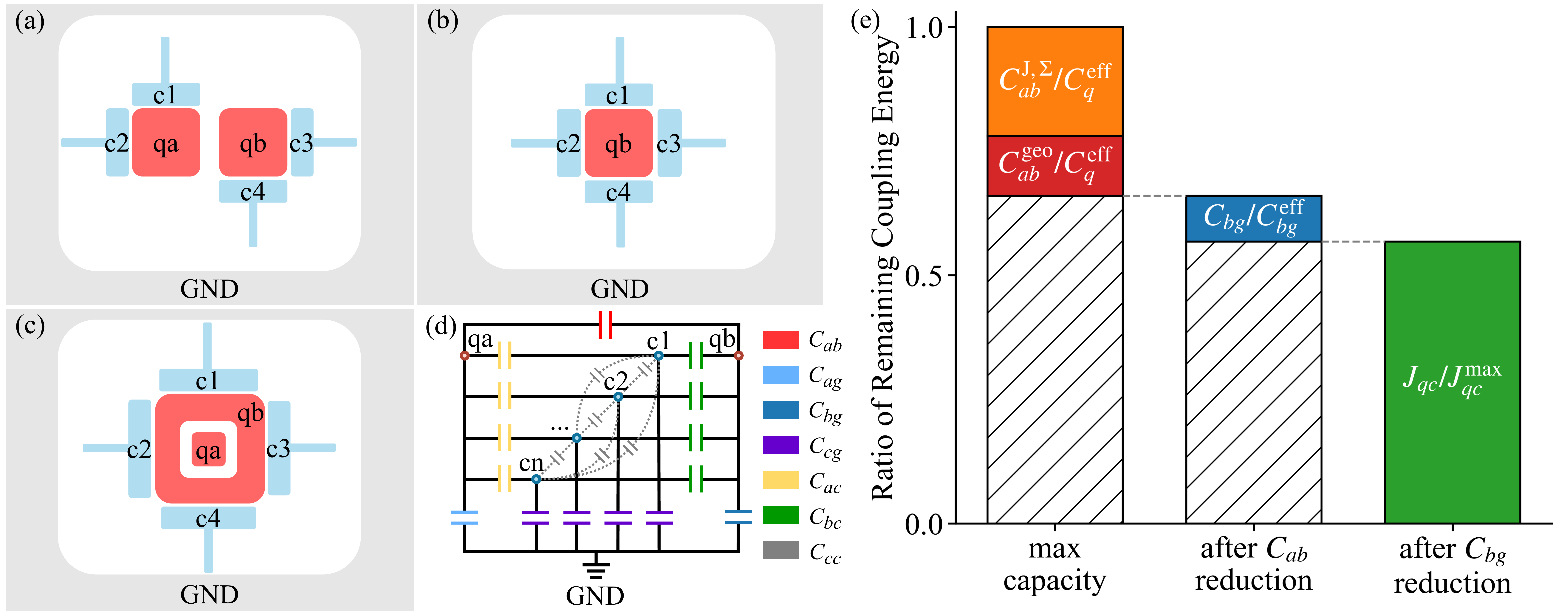}
\caption{Qubit–coupler coupling configurations, circuit model, and capacitive loading decomposition. Representative qubit–coupler coupling geometries: (a) Dumbbell floating qubit, with pads qa and  qb coupled by 2 coupler pads each. (b) Grounded qubit, with pad qb coupled by 4 coupler pads. (c) Concentric floating qubit, with pad qb coupled by 4 coupler pads.  (d) Equivalent circuit model of a floating qubit coupled by $X$ couplers. (e) Decomposition of the maximum achievable qubit–coupler capacitive coupling energy, $J_{qc}^{\max}/h = 4e^2/(XC_{cg})\eta_{qc}\eta_{cc}$. Orange and red: The reduction fraction  $C_{ab}/C_q^{\mathrm{eff}}$  induced by  $C_{ab}$, where  $C_{ab}$  includes both the geometric pad-to-pad capacitance,  $C_{ab}^{\text{geo}}$, and the total junction-induced parasitic capacitance (consisting of JJ and JJA), $C_{ab}^{\mathrm{J,\Sigma}}$. Blue: The reduction fraction  $C_{bg}/C_{bg}^{\mathrm{eff}}$  induced by the pad-to-ground capacitance,  $C_{bg}$, among the remaining $(1 - C_{ab}/C_q^{\mathrm{eff}})$  budget. Green: The fraction available for qubit–coupler coupling, $J_{qc}/J_{qc}^{\max} = (1 - C_{ab}/C_q^{\mathrm{eff}})(1 - C_{bg}/C_{bg}^{\mathrm{eff}})$.}
    \label{fig: schematic}
\end{figure*}

\section{Circuit Model and Theoretical Framework} 

\subsection{Minimal Model: A Grounded Qubit–Coupler System}
Understanding capacitive loading requires a quantitative relation between a qubit's capacitance budget and its achievable capacitive coupling to external circuit elements. We start with a minimal model consisting of a grounded qubit ($q$) capacitively coupled to a grounded coupler ($c$). Restricting to the kinetic-energy sector, the circuit Lagrangian in the flux basis is written as
$T=\dot{{\boldsymbol{\Phi}}}^{T}\mathbf{C}\dot{{\boldsymbol{\Phi}}}/2$, with ${\boldsymbol{\Phi}}^{T}=({\Phi}_q,{\Phi}_c)$ and
$\mathbf{C}=
\begin{pmatrix}
C_{qg}+C_{qc} & -C_{qc}\\
-C_{qc} & C_{cg}+C_{qc}
\end{pmatrix}$.
Here, $C_{qg}$ and $C_{cg}$ denote the direct qubit-to-ground and coupler-to-ground capacitances, respectively, while $C_{qc}$ represents the mutual capacitance between the qubit and coupler. Transforming into the charge basis gives $\hat{H}=\hat{\mathbf{Q}}^{T}\mathbf{C}^{-1}\hat{\mathbf{Q}}/2$, with $\hat{\mathbf{Q}}^{T}=(\hat{Q}_q,\hat{Q}_c)$. The resulting effective capacitances for the qubit and coupler are solved respectively.
\begin{equation}
C_{q}^{\mathrm{eff}}=C_{qg}+\frac{C_{cg}C_{qc}}{C_{cg}+C_{qc}},
\quad
C_{c}^{\mathrm{eff}}=C_{cg}+\frac{C_{qg}C_{qc}}{C_{qg}+C_{qc}}.
\end{equation}
The qubit--coupler interaction Hamiltonian takes the form $\hat{H}_{qc}=J_{qc}\hat{n}_q\hat{n}_c$, where $\hat{n}_q=\hat{Q}_q/(2e)$ and $\hat{n}_c=\hat{Q}_c/(2e)$ denote the Cooper-pair number operators of the qubit and coupler. The coupling strength is obtained as
\begin{equation}\label{Jqc_2mode}
J_{qc}=\frac{4e^2}{C_{cg}}\left(1-\frac{C_{qg}}{C_{q}^{\mathrm{eff}}}\right)
=\frac{4e^2}{C_{qg}}\left(1-\frac{C_{cg}}{C_{c}^{\mathrm{eff}}}\right),
\end{equation}
where $C_{qg}/C_{q}^{\mathrm{eff}}$ and $C_{cg}/C_{c}^{\mathrm{eff}}$ are the capacitance participation ratios associated with the qubit and coupler grounding channels. This result establishes an explicit quantitative relation between the capacitance budget and the effective interaction strength. Importantly, it shows that the coupling strength is determined by both the bare grounding capacitances and the fraction of the qubit capacitance budget consumed by its grounding channel. This motivates a participation-based description of capacitive loading in more highly connected
architectures.

\subsection{Generalization to Highly Connected Architectures: A Unifying Analytical Relation}

Next, we extend our analysis to highly connected 2D architectures. As illustrated in Fig.~\ref{fig: schematic}, the fluxonium qubit geometries are broadly classified into three configurations: dumbbell floating~\cite{Nguyen2019_fluxonium,Fei2025}, grounded~\cite{Najera-Santos2024_Grounded_Fluxonium}, and concentric floating~\cite{Hida2025_Concentric_Fluxonium}. To obtain a compact analytical result, we focus on a representative symmetric dumbbell floating qubit. In this configuration, each pad is capacitively coupled to $X$ identical grounded couplers, giving $N=2X$ total couplers per qubit. The conventional four-coupler connectivity of a two-dimensional grid corresponds to $X = 2$, as shown in Fig.~\ref{fig: schematic}(a).

Denoting the node fluxes of the qubit pads by $\Phi_a, \Phi_b$ and those of the attached couplers by $\Phi_{a_1}, \Phi_{a_2}, \Phi_{b_1}, \Phi_{b_2}$, the kinetic-energy contribution to the circuit Lagrangian is $T = \dot{\boldsymbol{\Phi}}^{T}\mathbf{C}\dot{\boldsymbol{\Phi}}/2$, where $\boldsymbol{\Phi}=(\Phi_a,\Phi_b,\Phi_{a_1},\Phi_{a_2},\Phi_{b_1},\Phi_{b_2})^T$. 
For simplification, we assume identical coupler geometries with uniform grounding and mutual capacitances.
Given the spatial and electromagnetic symmetry of the two pads, the capacitive parameters remain invariant under the pad index exchange $a \leftrightarrow b$. Retaining the intended capacitive couplings alongside the dominant parasitic paths, the resulting capacitance matrix $\mathbf{C}$ takes the explicit form
\begin{equation}
\mathbf{C}=\resizebox{0.88\columnwidth}{!}{$\begin{pmatrix}
C_q&-C_{ab}&-C_{qc}&-C_{qc}&-C_{Xqc}&-C_{Xqc}\\
-C_{ab}&C_q&-C_{Xqc}&-C_{Xqc}&-C_{qc}&-C_{qc}\\
-C_{qc}&-C_{Xqc}&C_c&-C''_{cc}&-C'_{cc2}&-C'_{cc1}\\
-C_{qc}&-C_{Xqc}&-C''_{cc}&C_c&-C'_{cc1}&-C'_{cc2}\\
-C_{Xqc}&-C_{qc}&-C'_{cc2}&-C'_{cc1}&C_c&-C''_{cc}\\
-C_{Xqc}&-C_{qc}&-C'_{cc1}&-C'_{cc2}&-C''_{cc}&C_c
\end{pmatrix}$},
\label{eq:capacitance_matrix}
\end{equation}
where $C_q=C_{bg}+C_{ab}+2C_{qc}+2C_{Xqc}$ and $C_c=C_{cg}+C_{qc}+C_{Xqc}+C'_{cc1}+C'_{cc2}+C''_{cc}$. Here, $C_{ab}$ is the inter-pad capacitance; $C_{qc}$ is the intended qubit--coupler capacitance; $C_{Xqc}$ couples a qubit pad to couplers attached to the opposite pad; $C'_{cc1}$ and $C'_{cc2}$ couple couplers attached to opposite qubit pads; and $C''_{cc}$ couples neighboring couplers attached to the same pad.

To isolate the qubit degree of freedom, we transform the qubit-pad node fluxes into differential and common modes, $\Phi_- = \Phi_a - \Phi_b$ and $\Phi_+ = \Phi_a + \Phi_b$. Following the standard circuit quantization procedure~\cite{Devoret1995_CQED, Blais2021_CQED}, the common mode $\Phi_+$ decouples from the potential energy and is eliminated as an uncoupled free variable.
This coordinate transformation yields a transformed capacitance matrix $\mathcal{C}$, whose inverse $\mathcal{C}^{-1}$ directly determines the Hamiltonian parameters. Generally, the diagonal elements of $\mathcal{C}^{-1}$ specify the charging energy of mode $i$ through
$E_C^i = \frac{e^2}{2} \mathcal{C}^{-1}_{ii}$
while the off-diagonal elements define the capacitive coupling strength between modes $i$ and $j$ through
$J_c^{(i,j)} = 4e^2 \mathcal{C}^{-1}_{ij} \quad (i \neq j)$.
Retaining only the degrees of freedom governing qubit--coupler interactions, the Hamiltonian takes the form 
\begin{equation}
\hat{H}
=
4E_C^q\hat{n}_q^2
+
\sum_{c}
\left(
4E_C^c\hat{n}_c^2
+
J_{qc}\hat{n}_q\hat{n}_c
\right),
\end{equation}
where $c\in\{a_1,a_2,b_1,b_2\}$ indexes the couplers. The effective charging energies are defined as $E_C^q = e^2 / (2C_q^{\mathrm{eff}})$ and $E_C^c = e^2 / (2C_c^{\mathrm{eff}})$, with $C_q^{\mathrm{eff}}$ and $C_c^{\mathrm{eff}}$ being the effective qubit and coupler capacitances, respectively.

\begin{table*}[!t]
\caption{Qubit effective capacitance $C_q^\mathrm{eff}$ and qubit–coupler coupling strength $J_{qc}$ for various configurations}
\label{tab:qc_config}
\centering
\begin{tabular}{cccc}
\hline\hline
Configuration & Dumbbell floating & Grounded & Concentric  floating \\

& Fig.~\ref{fig: schematic}(a)
& Fig.~\ref{fig: schematic}(b)
& Fig.~\ref{fig: schematic}(c)
\\
\hline

\multirow{1}{*}{$C^{\mathrm{eff}}_{q}$}

&
& 
$C_{ab}+\left[(C_{ag}^{\mathrm{eff}})^{-1}+(C_{bg}^{\mathrm{eff}})^{-1}\right]^{-1}
\vspace{0.1cm}

$\\

\hline
\vspace{0.1cm}
\multirow{1}{*}{$J_{qc}$}
&
&$\frac{4e^{2}}{XC_{cg}}\left(1-\frac{C_{ab}}{C_{q}^{\mathrm{eff}}}\right)\left(1-\frac{C_{bg}}{C_{bg}^{\mathrm{eff}}}\right)$
&
\\

\hline

\multirow{1}{*}
{$X$}
& 2
& 4
& 4 \\

{$C_{ab}$}
& $C_{ab}$
& 0
& $C_{ab}$ \\

{$C^{\mathrm{eff}}_{ag}$}
& $C_{ag}+2\left(C_{cg}^{-1}+C_{qc}^{-1}\right)^{-1}$
& N.A.
&  $C_{ag}$\\

$C^{\mathrm{eff}}_{bg}$
&$C_{bg}+2\left(C_{cg}^{-1}+C_{qc}^{-1}\right)^{-1}$
&$C_{bg}+4\left(C_{cg}^{-1}+C_{qc}^{-1}\right)^{-1}$
&$C_{bg}+4\left(C_{cg}^{-1}+C_{qc}^{-1}\right)^{-1}$
\vspace{0.1cm}
\\

\hline\hline
\end{tabular}
\end{table*}

By analytically evaluating the inverse matrix elements of $\mathcal{C}^{-1}$, we obtain a compact expression for the qubit–coupler coupling strength $J_{qc}$, given by Eq.~\eqref{eq: analytical_Jqc}. The detailed derivation is provided in Appendix~\ref{sec: symm_fqgc}, and its accuracy is validated against full electromagnetic simulation and  numerical circuit quantization post-processing in Appendix~\ref{sec: EM_1q4c}. 
\begin{equation}\label{eq: analytical_Jqc}
J_{qc}
=
\frac{4e^2}{XC_{cg}}
\eta_{qc}\eta_{cc}
\left(
1-\frac{C_{ab}}{C_q^{\mathrm{eff}}}
\right)
\left(
1-\frac{C_{bg}}{C_{bg}^{\mathrm{eff}}}
\right).
\end{equation}
Here, $C_{cg}$ denotes the bare coupler-to-ground capacitance, and $X$ is the number of couplers connected to a given qubit pad ($p$). The factor $\eta_{qc}=(1-r_1)/(1+r_1+4r_2)$ quantifies the impact of parasitic capacitances, where $r_1=C_{Xqc}/C_{qc}$ and $r_2=C_{Xqc}/(C_{cg}+2C_{cc})$.  $C_{Xqc}$ denotes the mutual capacitance between the qubit pad and couplers not directly connected to it. An additional suppression factor, $\eta_{cc}=(1+2C_{cc}/C_{cg})^{-1}$, accounts for coupler--coupler parasitic coupling, where $C_{cc}$ is the total capacitance between couplers connected to different qubit pads. 
Notably‌, the analytical result separates the coupling-budget into distinct physical channels: parasitic capacitance due to non-target elements, coupler--coupler cross-capacitance, inter-pad capacitance, and pad-to-ground capacitance. This decomposition provides a direct mapping between the capacitance matrix and experimentally relevant geometric design variables, while recovering Eq.~\eqref{Jqc_2mode} as the limiting case $X=1$, $\eta_{qc}=\eta_{cc}=1$, and $C_{ab}=0$.

As indicated by Eq.~\eqref{eq: analytical_Jqc}, the upper bound on the qubit–coupler coupling strength is set by  $4e^2/(XC_{cg})$. Previous study to mitigating capacitive loading~\cite{Ding2023_MIT_FTF,Rosenfeld2024,Peng2026,Schirk2026} primarily increase this upper bound by reducing $C_{cg}$, as captured by our analytical relation. Our finding further reveals that the allocation of the remaining capacitance among competing coupling and parasitic channels ultimately determines the fraction of this upper bound that can be realized. In realistic devices, this bound is further reduced by parasitic capacitances, such as  $C_{Xqc}$  and  $C_{cc}$. The available capacitive coupling energy,  $J_{qc}^{\rm max} = 4e^2\eta_{qc}\eta_{cc}/(XC_{cg})$, is then distributed among multiple coupling pathways (Fig.~\ref{fig: schematic}(e)). A central insight is that the resulting qubit–coupler coupling is governed by two crucial capacitance participation ratios: (i) the inter-pad participation,  $C_{ab}/C_q^{\rm eff}$, defined as the fraction of the qubit's effective capacitance associated with the inter-pad capacitive channel; and (ii) the qubit–ground participation,  $C_{bg}/C_{bg}^{\mathrm{eff}}$, defined as the fraction of the effective pad-to-ground capacitance contributed by the direct pad-to-ground path.

While the above derivation focuses explicitly on the dumbbell-floating configuration, the grounded and concentric-floating geometries can be treated within the same circuit-quantization framework, with their distinct geometries entering only through the corresponding capacitance matrices. The detailed analytical derivations for these configurations are presented in Appendices~\ref{sec:groundQ_groundC} and~\ref{sec:asyQ_groundC}, respectively. The resulting key parameters, including the qubit effective capacitance $C_q^{\mathrm{eff}}$, the effective pad-to-ground capacitance $C_{bg}^{\mathrm{eff}}$, and the coupling strength $J_{qc}$, are summarized in Table~\ref{tab:qc_config}. Remarkably, in the absence of selected parasitic capacitances, all three configurations reduce to the same analytical expressions for $C_q^{\mathrm{eff}}$ and $J_{qc}$, establishing a unified function structure and electrostatic description of capacitive loading across these distinct qubit configurations.

\begin{figure*}[t]
    \centering
    \includegraphics[width=0.81\linewidth]{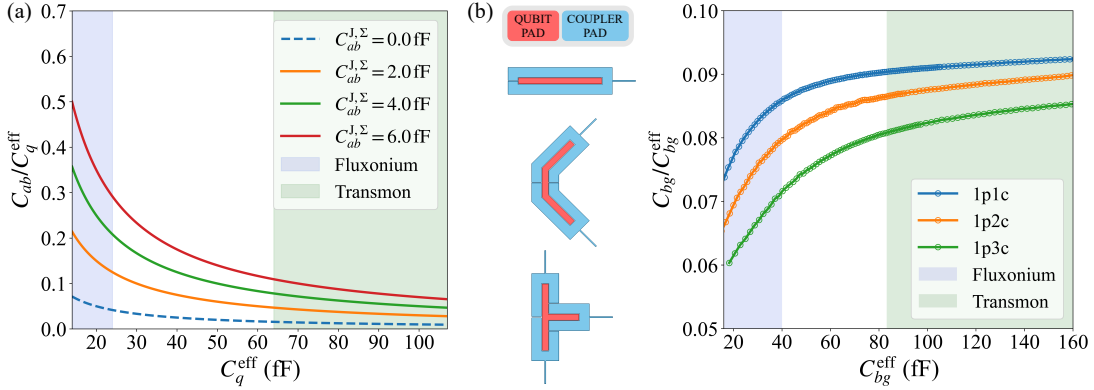}
    \caption{Scaling behavior of the two critical capacitance participation ratios governing capacitive loading. (a) Inter-pad capacitance participation ratio  $C_{ab}/C_q^{\rm eff}$  as a function of the qubit effective capacitance  $C_q^{\rm eff}$, with fixed geometric contribution $C_{ab}^{\rm geo}=1\,\text{fF}$. Increasing junction parasitic capacitance $C_{ab}^{\rm J,\Sigma}$ substantially increases $C_{ab}/C_q^{\rm eff}$ in fluxonium regime (blue), while having a much weaker effect in transmon regime (green) with larger capacitance budget.  (b) Electromagnetically optimized layouts for one-pad-one-coupler (1p1c), one-pad-two-coupler (1p2c), and one-pad-three-coupler (1p3c) configurations and the corresponding pad-to-ground participation ratio, $C_{bg}/C_{bg}^{\rm eff}$, as a function of the effective pad-to-ground capacitance $C_{bg}^{\rm eff}$.
    The optimized layouts maintain $C_{bg}/C_{bg}^{\rm eff}<0.1$ across different qubit connectivities for both fluxonium and transmon regimes.}
    \label{fig: em-eng}
\end{figure*}

\section{Mitigating Strategy and Design Principles} \label{sec:MS_DP}

\subsection{Mitigating Strategy}

The analytical framework obtained in the last section separates capacitive loading into distinct physical contributions, suggesting a hierarchical electromagnetic design strategy with two complementary objectives: (i) maximize the achievable upper bound $J_{qc}^{\rm max}$ by minimizing parasitic capacitance overhead, and (ii) preserve the available coupling budget by suppressing the residual loading factors associated with  $C_{ab}/C_q^{\rm eff}$  and  $C_{bg}/C_{bg}^{\rm eff}$. 

First, the upper bound of $J_{qc}$ is maximized by minimizing $C_{cg}$, $C_{Xqc}$, $C_{cc}$, and $X$. Reducing the coupler-to-ground capacitance  $C_{cg}$  directly enhances the upper bound~\cite{Peng2026_F_CapLd}, although its suppression is limited by irreducible contributions from coupling segments, finite qubit--qubit spacing and JJ parasitic capacitance. Further reduction of $C_{cg}$  generally corresponds to shorter coupling extensions and smaller qubit separation, at the cost of increased wiring complexity and potential microwave crosstalk. Unintended capacitance channels, $C_{Xqc}$ and $C_{cc}$, should likewise be minimized through spatial separation or grounded shielding of neighboring couplers. Additionally, capacitive loading should be distributed uniformly across both of the qubit pads. For a fixed qubit connectivity, such distribution reduces the per-pad loading factor $X$ and enhances the achievable  $J_{qc}$. In floating qubits, loading is naturally shared between two pads, yielding a twofold enhancement of the  $J_{qc}$  upper bound relative to grounded architectures with identical connectivity. Therefore, among three configurations considered, the dumbbell-shaped floating qubit is the most favorable for mitigating capacitive loading. 

Next, we consider the suppression of the loading factor $C_{ab}/C_q^{\mathrm{eff}}$, which arises from the mutual capacitance between the qubit pads $a$ and $b$. The total capacitance is defined as $C_{ab} = C_{ab}^{\mathrm{geo}} + C_{ab}^{\mathrm{J,\Sigma}}$, where $C_{ab}^{\mathrm{geo}}$ represents the direct bare geometric capacitance between the pads, and $C_{ab}^{\mathrm{J,\Sigma}} = C_{ab}^{\mathrm{JJ}} + C_{ab}^{\mathrm{JJA}}$ is the total junction-induced parasitic capacitance. Here, $C_{ab}^{\mathrm{JJ}}$ denotes the parasitic capacitance induced by the JJ, while $C_{ab}^{\mathrm{JJA}}$ represents the contribution from the JJA, which is modeled here as an approximately constant background \footnote{For transmon qubits, the array contribution naturally vanishes ($C_{ab}^{\mathrm{JJA}} = 0$), whereas it is a loading channel for fluxonium qubits.}. The geometric contribution $C_{ab}^\mathrm{geo}$ can be reduced by widening pad separation and using surrounding coupler structures for electrostatic shielding. By contrast, $C_{ab}^{\mathrm{JJ}}$, which scales with junction area, is determined by $E_J$ and fabrication-dependent critical current density. In practice, the minimum physical junction size (typically $\sim 100\ \times 100\,\text{nm}^2$) is constrained by lithographic resolution, preventing further area reduction~\cite{Hertzberg2021_laser_annealing_JJ}. Hence, junction parasitic capacitance constitutes a non-negligible fraction of $C_{ab}$ and sets a process-dependent lower bound for loading suppression. For one typical junction process, e.g., $C_{ab}^{\mathrm{JJ}} \approx 0.8\,\text{fF/GHz} \times E_J/h$, yielding $C_{ab}^{\mathrm{JJ}} \sim 4\,\text{fF for a representative } E_J/h = 5\,\text{GHz}$. This effect is particularly crucial in fluxonium qubits, which typically possess a relatively small capacitance budget, $C_q^{\mathrm{eff}} \sim 14\text{--}24 \,\textrm{fF}$. As revealed in Fig.~\ref{fig: em-eng}(a), $C_{ab}/C_q^{\mathrm{eff}}$ is intrinsically larger in fluxonium regime, making capacitive loading significantly more severe in fluxonium architectures. Reducing junction parasitic capacitance~\cite{Watanabe2003_smallCJJ,Shapovalov2020_lowCJJ} through thinner oxide barriers or higher critical current densities is therefore a key mitigation strategy, as it directly reduce the junction area required to achieve a given $E_J$, thereby suppressing $C_{ab}^{\mathrm{JJ}}$.

Finally, we consider the suppression of the loading factor  $C_{bg}/C_{bg}^{\mathrm{eff}}$ through constructing an electromagnetic model with three distinct qubit pad connectivities using a half-qubit geometry, as shown in Fig.~\ref{fig: em-eng}(b) (see details in Appendix~\ref{sec: EM_general_setup} for the general EM simulation setup adopted throughout this work). For the dumbbell floating qubit with symmetric coupler loading, the qubit connectivity $N = 2X$ serves as a key metric for hardware-efficient quantum error correction, with representative connectivities $N=2,4,6$ compatible with repetition-code~\cite{chen2021_repetition}, surface-code~\cite{GoogleQAI_2025}, and qLDPC-code~\cite{Wang2026_qLDPC}, respectively. 
Note that these configurations remove  $C_{Xqc}$,  $C_{ab}$  and  $C_{cc}$, therefore $C_{bg}^{\mathrm{eff}}$ is reduced to  $C_{bg} + \sum_{i=1}^{X} (C_{bc,i}^{-1} + C_{cg,i}^{-1})^{-1}$, with $X$ coupler pads loaded on one qubit pad (`p'). For symmetric floating-qubit with symmetric coupler loading, $C_{ag}^\mathrm{eff}=C_{bg}^\mathrm{eff}$, yielding $C_{q}^{\mathrm{eff}}=C_{ab}+C_{bg}^{\mathrm{eff}}/2$. Therefore, $C_{bg}^\mathrm{eff}$ serves as an equivalent metric for quantifying the capacitance budget as well in this special case. Fig.~\ref{fig: em-eng}(b) summarizes three main findings.
First, systematic geometric optimization based on electromagnetic simulations suppresses $C_{bg}/C_{bg}^{\mathrm{eff}}$ below 0.1 for all three different connectivity configurations. The lower bound of $C_{bg}/C_{bg}^{\mathrm{eff}}$ is set by the minimum qubit–coupler gap permitted by the fabrication process. In present designs, this gap is fixed at $2\,\mu\mathrm{m}$; reducing it further would lead to even smaller values of $C_{bg}/C_{bg}^\mathrm{eff}$ (see details in Appendix~\ref{sec:r_bg}).
Second, within capacitance regimes relevant to fluxonium and transmon qubits,  $C_{bg}/C_{bg}^{\mathrm{eff}}$  remains largely insensitive to the qubit's capacitance budget. 
Third,  $C_{bg}/C_{bg}^{\mathrm{eff}}$ is nearly independent of qubit connectivity $X$; the observed differences across configurations arise primarily from geometry-dependent electromagnetic constraints in practical layouts. 

\begin{table*}[t]
\caption{Practical design principles for mitigating capacitive loading}
\label{tab:design_rules}
\centering

\renewcommand{\arraystretch}{1.2}

\begin{tabular}{lcc}
\hline
\hline

\textbf{\makecell[l]{Design Rules}}
&\textbf{Quantities}
&\textbf{Approaches}
\\

\midrule

\makecell[l]{1. Distribute capacitive loading uniformly}
&
$\downarrow X$
&
\makecell[l]{
(i) Distribute couplers symmetrically on floating qubit pads \\
(ii) Design equal coupler-to-ground capacitances
}
\\

\midrule

\makecell[l]{2. Suppress parasitic pairwise coupling}
&
$\downarrow C_{Xqc},\ \downarrow C_{cc}$
&
\makecell[l]{
(i) Increase pad spacing \\
(ii) Spread couplers with equally distributed angles \\
(iii) Improve geometric screening by coupled qubit pad
}
\\

\midrule

\makecell[l]{3. Minimize coupler grounding capacitance}
&
$\downarrow C_{cg}$
&
\makecell[l]{
(i) Design lower $E_J^c$ \\
(ii) Decrease junction area and oxidation \\
(iii) Increase ground padding
}
\\

\midrule

\makecell[l]{4. Minimize direct inter-island capacitance}
&
$\downarrow C_{ab}/C_q^\mathrm{eff}$
&
\makecell[l]{
(i) Design lower $E_J^q$ \\
(ii) Decrease junction area and oxidation \\
(iii) Increase pad spacing
}
\\

\midrule

\makecell[l]{5. Suppress qubit parasitic grounding paths \\ ~~~~and enhance targeted pairwise coupling}
&
$\downarrow C_{bg}/C_{bg}^\text{eff}$
&
\makecell[l]{
(i) Increase geometric confinement from coupler pad\\
(ii) Increase coupling section area \\
(iii) Decrease pads coupling sections spacing \\
(iv) Increase pads parallel length
}
\\

\hline
\hline
\end{tabular}
\end{table*}

\subsection{Physical Insights and Design Principles}

 The above theoretical and electromagnetic analysis establishes a unifying physical picture of capacitive loading. The achievable qubit--coupler interaction is therefore not governed by the absolute capacitance values, but by how the finite qubit capacitance budget is distributed among competing capacitance channels, as quantified by two capacitance participation ratios in Eq.~\eqref{eq: analytical_Jqc}. The framework reveals two distinct optimization pathways. Within the fluxonium regime, parasitic capacitances induced by the JJs and JJAs constitute a non-negligible fraction of this limited budget, making the inter-pad participation $C_{ab}/C_q^{\rm eff}$ an important constraint that ultimately calls for improvements in junction fabrication process, while the pad-to-ground participation $C_{bg}/C_{bg}^{\rm eff}$ can be systematically reduced via electromagnetic layout optimization. 

Although this work is motivated by investigating the tradeoff between the limited capacitance budget of fluxonium qubits and their achievable coupling to external circuit elements, the underlying physical mechanism is not specific to fluxonium. As illustrated in Fig.~\ref{fig: em-eng}, the same capacitance-partitioning mechanism applies across the fluxonium (blue) and transmon (green) regimes. As $C_q^{\rm eff}$ increases toward the transmon regime, the capacitances associated with the same junctions constitute a progressively smaller fraction of the total qubit capacitance, thereby reducing $C_{ab}/C_q^{\rm eff}$ and providing greater headroom for $C_{bg}/C_{bg}^{\rm eff}$. This additional capacitance headroom is particularly relevant to tunable-coupler architectures \cite{Yan2018}, which are widely employed in high-performance scalable superconducting processors, e.g., \textit{Sycamore} and the \textit{Zuchongzhi} series~\cite{GoogleQAI_2019,Zuchongzhi2.1_2021,Zuchongzhi3.2_2025}. In such architectures, a larger qubit–coupler coupling $g_{qc}$ enables a stronger effective qubit–qubit interaction, $g_{qq}^{\mathrm{eff}}\sim g_{qc}^{2}/\Delta_{qc}$, while maintaining a dispersive regime with $g_{qc}/\Delta_{qc}\ll1$~\cite{GoogleTunableCouplerPatent,Li2023_QCQ}. The resulting combination of strong effective interaction and weak qubit–coupler hybridization provides a route to faster two-qubit gates while preserving qubit localization and suppressing coupler-induced perturbations. Thus, suppressing capacitive loading is not only important for fluxonium processors but also provides a general design strategy for strong coupling strength within transmon-based architectures.

These implications motivate a general formulation of practical design principles for mitigating capacitive loading, as summarized in Table~\ref{tab:design_rules}. At electromagnetic and layout levels, the objective is to allocate the finite capacitance budget preferentially to the intended coupling channels while suppressing parasitic loading. This becomes increasingly important in scaled-up processors with degree-4 connectivity and beyond, where multiple coupling paths must coexist within a finite physical layout and inevitably compete with parasitic capacitance. 
Accordingly, the design principles span three interconnected levels: circuit topology, physical geometry, and device fabrication. Topologically, the circuit architecture determines the distribution of capacitive loading, favoring balanced connectivity and symmetry to reduce $X$. Geometrically, the spatial arrangement of pads and coupling sections shapes the electric-field distribution and determines the magnitude of both intended and parasitic capacitances, enabling suppression of $C_{Xqc}$, $C_{cc}$,  effective pad-to-ground loading while concentrating capacitance into the intended qubit--coupler coupling channel. Regarding fabrication, junction process parameters provide a direct handle for reducing the intrinsic capacitances associated with the qubit and coupler, thereby mitigating $C_{ab}$ and $C_{cg}$. Together, we establish a practical framework for translating participation-based physical criteria into concrete layout strategies for scalable superconducting processors.

\begin{figure*}[t]
    \centering
    \includegraphics[width=0.67\linewidth]{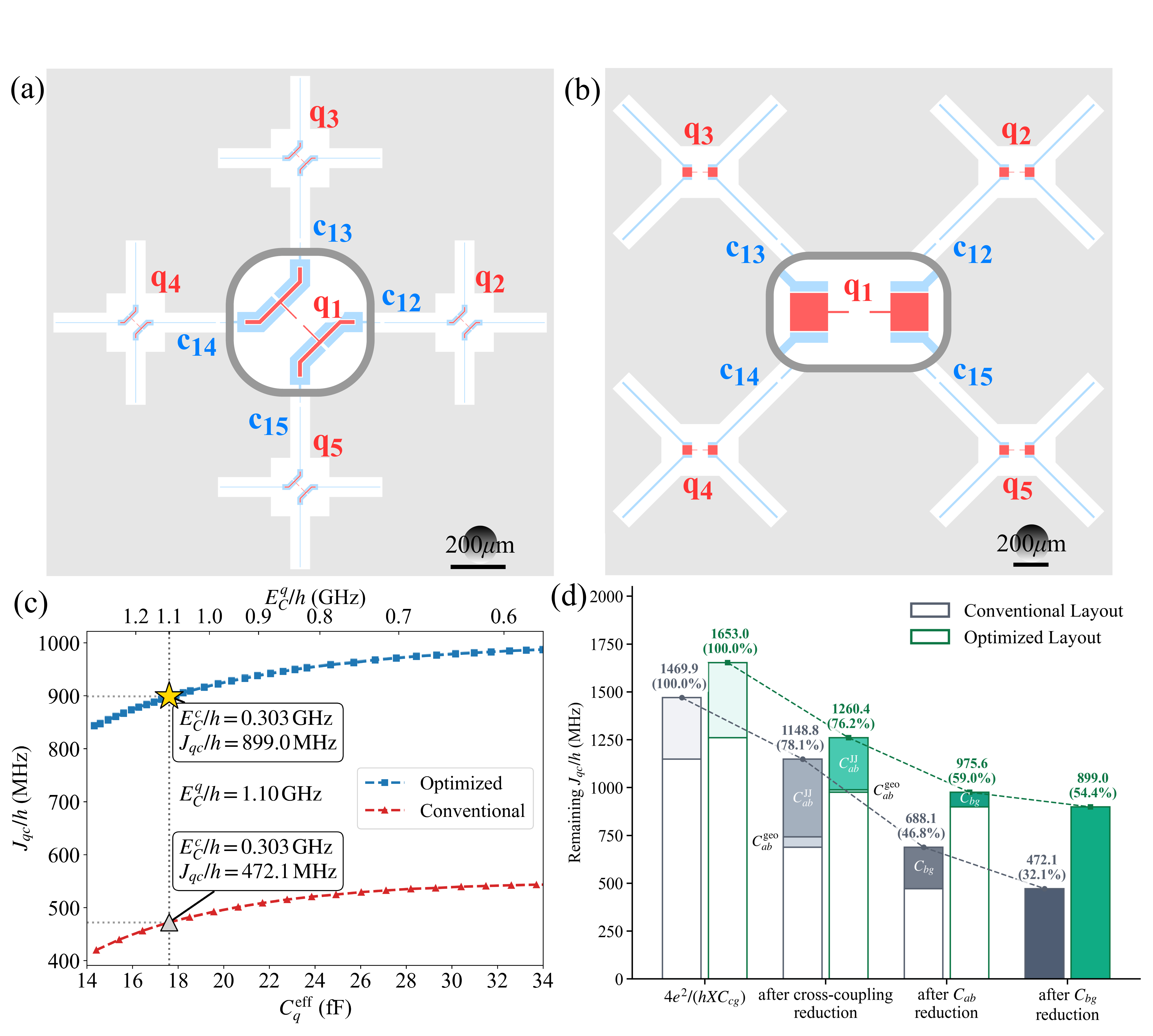}
    \caption{Demonstration of the proposed design principles in a scalable fluxonium processor.
(a) Schematic layout of the optimized 5Q4C scalable unit cell in a 2D grid fluxonium processor, featuring double-transmon couplers (DTCs) and implementing the capacitive loading mitigation strategy. $c_{ij}$ denotes the DTC connecting qubits $q_i$ and $q_j$.
(b) Schematic layout of the conventional 5Q4C unit cell for comparison, consisting of five square-shaped fluxonium qubits and four DTCs without geometric loading mitigation.
(c) Achievable qubit--coupler coupling strength, $J_{qc}$, for the optimized and conventional layouts as a function of the qubit effective capacitance, $C_q^\text{eff}$. The optimized design nearly doubles the achievable coupling while maintaining identical qubit and coupler charging energies.
(d) Cascade decomposition and reduction of achievable qubit–coupler coupling strength $J_{qc}$ between optimized (green) and conventional (grey) designs under identical $E_C^q$ and $E_C^c$. Starting from the theoretical upper bound ($4e^2 / hXC_{cg}$), sequential capacitive loading from $\eta_{qc}$, $\eta_{cc}$, $C_{ab}$ ($C_{ab}^{\mathrm{geo}} + C_{ab}^{\mathrm{J,\Sigma}}$), and $C_{bg}$ quantifies the budget erosion. Annotations and dashed lines show that the optimized design effectively suppresses this capacitive loading attenuation.}
    \label{fig: application_5q4c}
\end{figure*}

\section{Applications and Discussion} 
\subsection{A Scalable Unit Cell for 2D Grid Architectures
}\label{sec:5Q4C} 
Following the established design principles, we demonstrate their practical application in a $5$-qubit, $4$-coupler ($5\text{q}4\text{c}$) minimal scalable unit cell for 2D fluxonium grid architectures, as shown in Fig.~\ref{fig: application_5q4c}. For simplification, the electromagnetic simulations are performed with 2D planar layouts throughout this work. In practical implementations, flip-chip or multilayer architectures may introduce additional capacitive loading.
In the optimized layout, i.e., Fig.~\ref{fig: application_5q4c}(a), the fluxonium qubits inherit the wing-shaped two-arm geometry, where each arm serves as an independent coupling port. We pair these qubits with a double-transmon coupler (DTC) architecture~\cite{Chan2026_F_DTC, Goto2022_DTC, Li2024_DTC, Tiwari2026_DTC}, which intrinsically decouples capacitive and inductive channels to strongly suppress microwave, flux, and spectator-induced crosstalk while accommodating larger qubit spacing for control wiring. To maximize geometric shielding and suppress capacitive loading, each qubit arm and approximately half of its connecting segment are enclosed by the corresponding DTC coupling segment.
For comparison, we also design a conventional $5\text{q}4\text{c}$ unit cell (Fig.~\ref{fig: application_5q4c}(b)) that shares the same $1200\,\mu\text{m}$ qubit separation and $2\,\mu\text{m}$ qubit–coupler gap, as well as identical target charging energies ($E_C^q, E_C^c$) and Josephson energies ($E_J$), but employs a standard square-shaped fluxonium geometry with parallel hook-shaped coupler electrodes.

\begin{figure*}[t]
    \centering
    \includegraphics[width=0.71\linewidth]{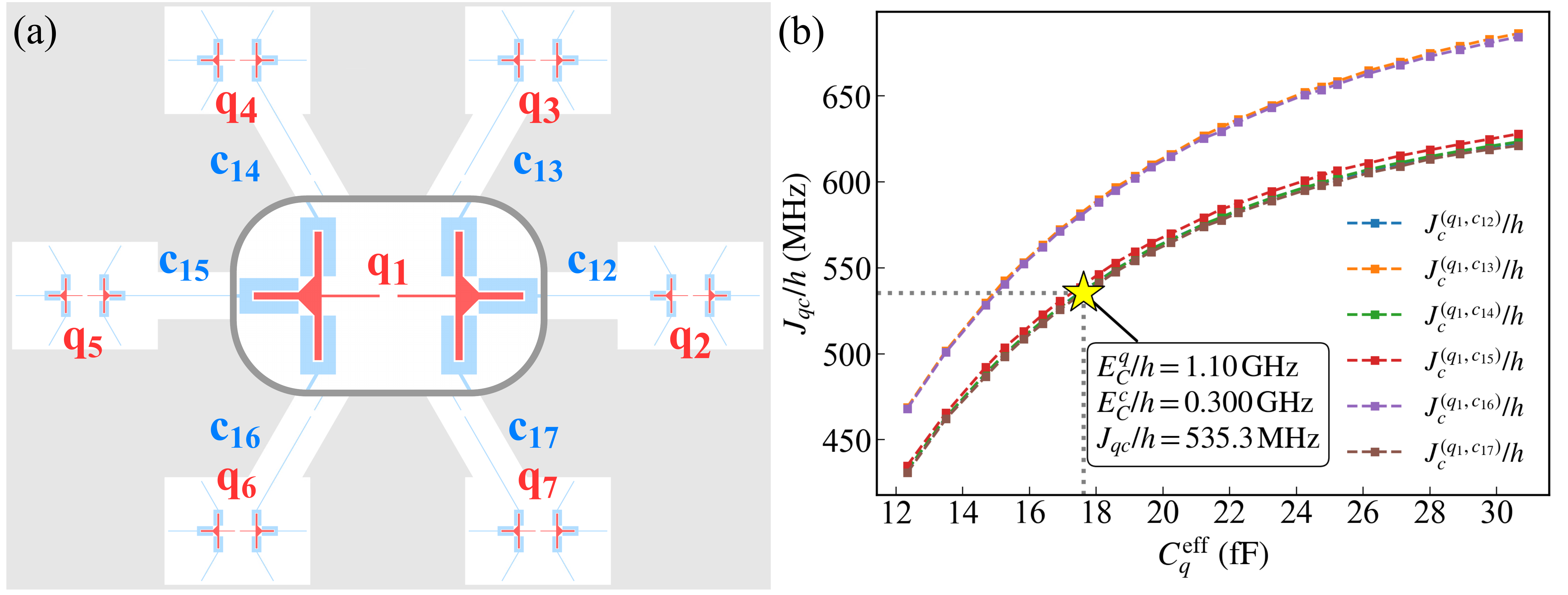}
    \caption{(a) Layout of 7q6c (7 qubits interconnected by 6 DTC) high-connectivity fluxonium processor designed using the proposed capacitive loading mitigation strategy. (b) Achievable qubit–coupler coupling strength $J_{qc}$ as a function of the effective qubit capacitance $C_q^\text{eff}$. At the target parameters $[E_C^q,E_C^c]/h=[1.1,0.3]\,\mathrm{GHz}$, all qubit–coupler pairs achieve $J_{qc}/h>535~\mathrm{MHz}$. Here, $J_c^{(q_i,c_{ij})}$ denotes the capacitive coupling strength between qubit $q_i$ and transmon coupler $c_{ij}$ connecting qubits $q_i$ and $q_j$.}
    \label{fig:7q6c}
\end{figure*}
By combining electromagnetic simulations with circuit-QED parameter extraction, we compare the achievable qubit–coupler coupling strength $J_{qc}$ for the two layouts respectively as a function of the effective qubit capacitance $C_q^\text{eff}$, as depicted in Fig.~\ref{fig: application_5q4c}(c). While the conventional layout limits the coupling strength to $J_{qc}/h \approx 470\,\mathrm{MHz}$, the optimized geometry nearly doubles this value, approaching $900\,\mathrm{MHz}$. This pronounced enhancement stems from two complementary mitigation strategies: 
i) Geometric shielding via the nearly full-enclosure wrapping structure, which drastically suppresses the pad-to-ground capacitance ratio $C_{bg}/C_{bg}^\text{eff}$ compared to the parallel coupling configuration; 
ii) Fabrication process optimization adopting a lower junction-specific parasitic capacitance of $(0.6 E_J / (h \cdot \mathrm{GHz}) + 1)\,\mathrm{fF}$, compared to $(0.8 E_J / (h \cdot \mathrm{GHz}) + 2.5)\,\mathrm{fF}$ in the conventional process, which suppresses both $C_{cg}$ and $C_{ab}$ associated with the Josephson junctions and Josephson junction arrays.
Importantly, this lower-capacitance baseline is achievable with state-of-the-art Josephson-junction fabrication techniques, thus representing a realistic engineering target \cite{Shapovalov2020_lowCJJ}. A practical trade-off, however, is that Josephson junctions with substantially different Josephson energies may require distinct oxidation conditions to maintain this suppressed parasitic capacitance.

To quantitatively dissect the physical origin of this enhancement, Fig.~\ref{fig: application_5q4c}(d) presents a step-by-step capacitive budget decomposition starting from the theoretical upper bound $4e^2 / (h X C_{cg})$. The cascade reduction quantifies how the capacitive budget is progressively reduced by the cross-capacitance factor ($\eta_{qc}$ and $\eta_{cc}$), inter-pad loading ($C_{ab} = C_{ab}^{\mathrm{geo}} + C_{ab}^{\mathrm{J,\Sigma}}$), and pad-to-ground shunting ($C_{bg}$). While both layouts experience similar initial attenuation from cross-capacitance, the conventional design loses a severe portion of its remaining budget to ground capacitance. In contrast, our optimized design rules effectively suppress the $C_{bg}$ loss stage, preserving a significantly higher fraction of the intrinsic coupling. Key parameter comparisons between the two designs are summarized in Appendix~\ref{subsec:comparison}.

Beyond the isolated qubit--coupler subsystem, practical implementations incur capacitive loading across two scales. Locally, auxiliary readout resonators and control lines introduce parasitic shunt capacitance that suppresses the achievable $J_{qc}$. Globally, embedding the sub-cell into a full lattice connect its nodes to surrounding circuit elements, inducing parasitic stray couplings with distant components and hence more capacitive loading.

\subsection{Generalization to High-Connectivity Architectures 
}\label{sec:7Q6C}

Beyond the connectivity-four configuration considered above, we further show that through applying the same capacitive loading mitigation strategies, large qubit--coupler coupling remains achievable in architectures with higher qubit connectivity. Optimizing the qubit-pad geometry and coupler placement enables unit cells supporting six nearest-neighbor couplings. Such architectures are compatible with lattice geometries for hardware-efficient quantum error-correction codes, such as the color code and qLDPC code~\cite{Chamberland2020_IBMColorCode, Breuckmann2021_QLDPC}. Fig.~\ref{fig:7q6c}(a) illustrates the extension of the proposed capacitive loading mitigation strategy to a six-neighbor fluxonium architecture. Following the design principles established in Sec.~\ref{sec:MS_DP}, a dumbbell floating fluxonium qubit is coupled to six neighboring qubits through double-transmon couplers (DTCs), realizing a hexagonal connectivity. The qubit adopts a three-branched T-shaped geometry, with each orthogonal arm serving as an independent coupling port enclosed by adjacent couplers. Increasing the qubit's arm length enlarges the coupling interface, enabling control of qubit self-capacitance and qubit–coupler mutual capacitance. 
Sweeping the arm length yields different combinations of $E_C^q$, $E_C^c$, and $J_{qc}$. Here, we target the same device parameters as in Sec.~\ref{sec:5Q4C}. The optimized design preserves the targeted charging energies while achieving a minimum qubit–coupler coupling strength of $J_{qc}/h=535.3~\mathrm{MHz}$ across all six coupling channels, as shown in Fig.~\ref{fig:7q6c}(b), which confirms that the proposed capacitive loading mitigation strategy can also be applied to higher-connectivity 2D fluxonium architectures.

\subsection{Fast and Robust Two-Qubit Gates Enabled by Suppressed Capacitive Loading}

To quantify the impact of suppressed capacitive loading and enhanced qubit–coupler interactions, we statistically benchmark MAP gate performance across an ensemble of 180 interacting two-qubit subsystems, corresponding to the nearest-neighbor coupling links for two-qubit gate of a $10\times10$ grid fluxonium processor architecture. The system Hamiltonian is provided in Appendix~\ref{subsec:System_Hamiltonian}. 
The capacitance parameters, and hence the corresponding charging energies $E_C$, are extracted from the optimized unit-cell layouts discussed above. Combined with the JJ parameters, the resulting circuit parameters are $[E_C^q,E_J^q,E_L^q]/h=[1.1,4.65,0.65]~\text{GHz}$ for the fluxonium qubit and $[E_C^{c_a},E_C^{c_b},E_J^{c_a},E_J^{c_b},E_J^{c_{ab}}]/h=[0.3,0.3,15,15,3.5]~\text{GHz}$ for the DTC, where $c_m$ ($m=a,b$) denotes transmon $a$ or $b$ of the DTC, and $E_J^{c_{ab}}$ denotes the Josephson energy of the junction connecting $c_a$ and $c_b$. Note that these parameters are determined following the parameter-design workflow introduced in Ref.~\cite{Chan2026_F_DTC}.
To assess the robustness of MAP gates against fabrication-induced parameter variations \cite{Hertzberg2021_laser_annealing_JJ, Pishchimova2023_JJ_Variation, Kennedy2025_JJ_Variation}, which become increasingly relevant in large-scale processors, we evaluate the subsystem performance under variations of $E_C$, $E_J$, and $E_L$ sampled around their nominal design values~\footnote{Variations in the nanoscale junction area and tunnel-barrier thickness lead to fluctuations in the critical current, thereby modifying $E_J$ and $E_L$. The same fabrication variations also alter the parasitic junction capacitance, resulting in deviations of $E_C$.}. Specifically, each parameter is independently sampled from a truncated Gaussian distribution centered at its nominal value, with a standard deviation of $\sigma = 5\%$ and symmetric truncation bounds at $\pm\sigma$. Using these perturbed parameter sets for the 180 subsystems, we evaluate the MAP gate infidelity within a noise-free model and determine the minimum gate time $t_{\rm gate}^*$ required to achieve a coherent leakage-induced infidelity below $10^{-3}$. 
It is worth pointing out that relaxation-induced errors are neglectable on the MAP gate timescale considered here. In particular, the two-qubit MAP gate is mediated by higher-energy states $\ket{\alpha}$ of the composite system, with major contributions from the $\ket{2}$ states of the fluxonium qubits. Relaxation from non-computational state $\ket{\alpha}$ to the computational subspace is characterized by the energy-relaxation time $T_{1,\alpha}$, which gives an estimated infidelity $1-F_{\rm relax}\approx{t_{\rm gate}}/{(8T_{1,\alpha})}$ in the limit $t_{\rm gate}\ll T_{1,\alpha}$~\cite{Ding2023_MIT_FTF}. 
A higher-level relaxation time of $T_{1,\alpha} \sim 5\,\mu\text{s}$~\cite{Ding2023_MIT_FTF, Somoroff2023} is sufficient to keep the decoherence-induced infidelity below $10^{-3}$ given $t_{\rm gate}\leq40\,\mathrm{ns}$. We can therefore focus on the coherent-leakage-limited gate performance, with further technical details provided in Appendix~\ref{subsec:Leakage_Analysis} and \ref{subsec:Gate_Fidelity}.

\begin{figure}[h]
    \centering
    \includegraphics[width=0.8\linewidth]{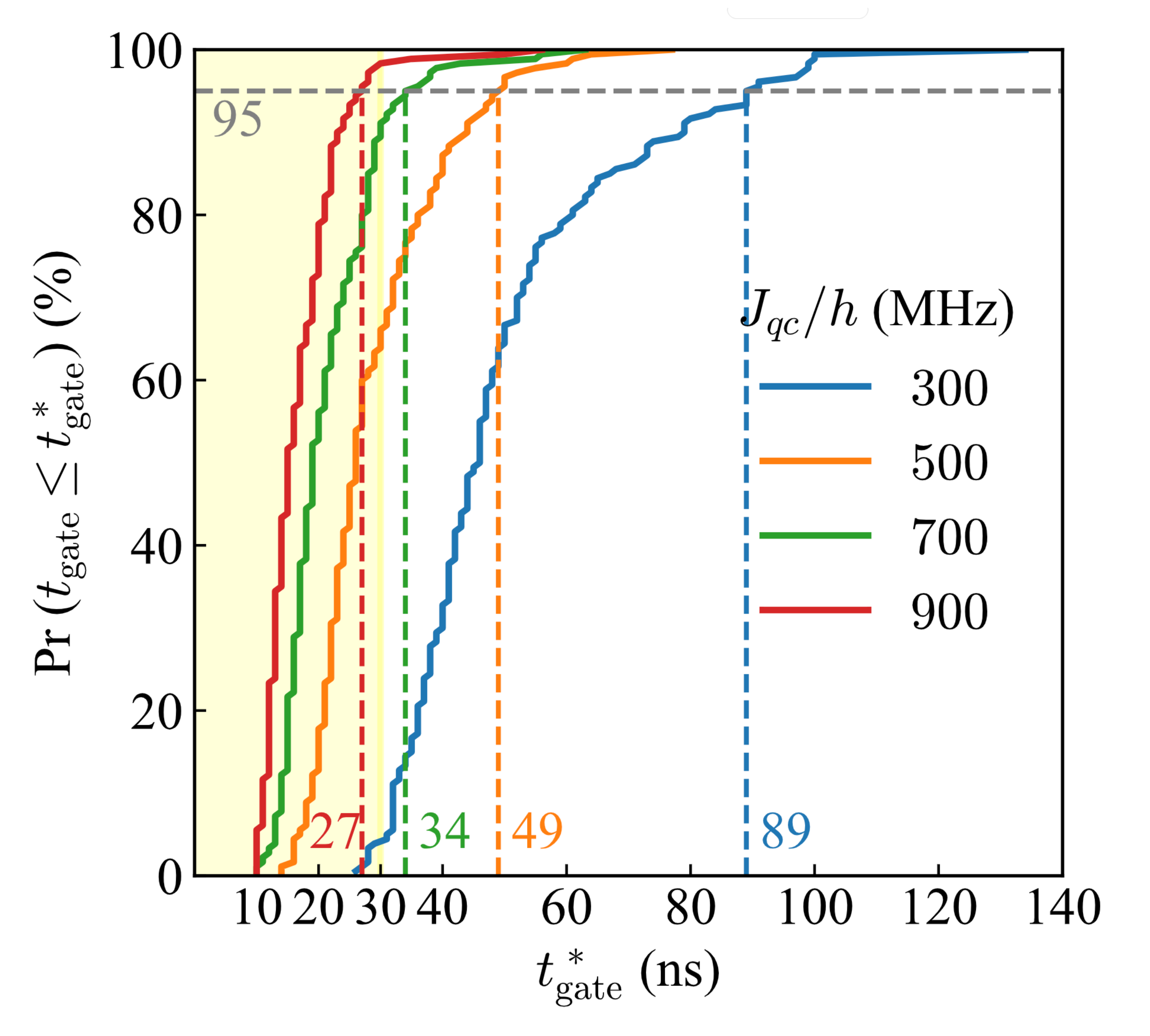}
    \caption{Impact of enhanced qubit–coupler coupling on MAP gate performance under 5\% fabrication-induced parameter variations. The cumulative probability of achieving a leakage-induced gate infidelity below $10^{-3}$ is shown as a function of the optimized gate time $t_{\rm gate}^*$ for $J_{qc}/h =300, 500, 700, 900\,\text{MHz}$. Statistics are obtained from 180 interacting qubit pairs. Stronger qubit–coupler coupling systematically shortens the achievable gate time and improves robustness against fabrication-induced parameter variations.}
    \label{fig: MAP_gate_time_robustness}
\end{figure}

Fig.~\ref{fig: MAP_gate_time_robustness} summarizes the cumulative distribution of $t_{\rm gate}^*$ for different qubit–coupler coupling strengths.  As $J_{qc}/h$ increases from 300 to $900\,\text{MHz}$, the 95th-percentile gate time decreases from 89 to 27 ns, with the corresponding distributions systematically shifting toward shorter gate times.  Within the noise-free leakage model, and considering the realistic fabrication variation, we find that stronger qubit--coupler coupling enables shorter MAP gates with reduced sensitivity to the modeled parameter variations. These results demonstrate that suppressing capacitive loading facilitates fast and robust two-qubit gates in highly connected 2D fluxonium processors.

\section{Conclusion} 
We establish a quantitative theory of capacitive loading in superconducting circuits through deriving a simple analytical relation between the finite qubit capacitance budget and the achievable qubit–coupler interaction strength. This framework reveals that capacitive loading is governed by two capacitance participation ratios with fundamentally different physical origins: the pad-to-ground participation ratio can be  suppressed through electromagnetic optimization, whereas the inter-pad participation ratio is ultimately dominated by parasitic capacitances associated with the Josephson junctions and Josephson junction arrays. 
Guided by these insights, we formulated and applied practical design principles to substantially increase the qubit–coupler coupling strength while preserving the target charging energies, and showed that stronger coupling enables shorter MAP-gate times within the considered coherent-leakage and fabrication-variation model. These results establish capacitive loading as an engineerable constraint rather than an unavoidable barrier within the class of highly connected fluxonium architectures. Beyond the fluxonium regime, the same capacitance-budget framework can be applied to transmon-based coupler architectures, where the increased capacitance headroom can be leveraged to realize stronger qubit–coupler coupling while maintaining dispersive operation, thereby opening a route to simultaneously faster two-qubit gates and reduced qubit–coupler hybridization for improved qubit localization.

While mitigating capacitive loading provides substantial benefits, it may introduce two potential trade-offs associated with the enclosing capacitor geometry and enhanced qubit–coupler coupling: (i) increased edge dielectric loss arising from a larger surface participation ratio due to the extended perimeter; and (ii) microwave crosstalk mediated by stronger capacitive interactions through the circuit network. 
For fluxonium qubits, both trade-offs can be mitigated with the present architecture. Geometry-induced dielectric loss is expected to remain secondary to the dominant decay channel associated with two-level systems in Josephson junction array oxide tunnel barriers, which has been discussed and demonstrated in Ref.~\cite{Zhuang_2026_Fluxonium_Loss}. Consequently, we can relax the geometric constraints on the pad to optimize coupling strengths, without incurring a significant penalty from surface dielectric loss. Meanwhile, enhanced qubit--coupler coupling does not necessarily imply stronger microwave crosstalk. DTC-based coupling schemes intrinsically suppress static capacitive mediation through the qubit--coupler network by interrupting the continuous capacitive coupling path with two inductively coupled transmons, in contrast to the direct capacitive mediation through a single transmon or resonator. 
Looking ahead, a comprehensive device-level evaluation will ultimately be needed to quantify these effects and establish whether they impose meaningful constraints on the experimentally achievable performance of highly connected fluxonium processors.

\appendix
\section{Derivation of Analytical Relation Between Qubit Capacitance and Qubit–Coupler Coupling} \label{sec:analytical_derivation}
This Appendix provides detailed derivations of the analytical expressions summarized in Table~\ref{tab:qc_config}, which characterize the effective qubit capacitance and qubit–coupler coupling strength. Section~\ref{sec: symm_fqgc} derives the corresponding expressions for the dumbbell floating geometry, explicitly incorporating realistic parasitic capacitances. Sections~\ref{sec:groundQ_groundC} and \ref{sec:asyQ_groundC} extend the analysis to the grounded and concentric floating geometries, respectively.
\subsection{Dumbbell floating qubit with grounded couplers}\label{sec: symm_fqgc}

We derive an analytical relation between the qubit–coupler coupling strength $J_{qc}$ and the effective qubit capacitance $C^\mathrm{eff}_q$. Particularly, we consider a dumbbell floating qubit coupled to $N=2X$ grounded couplers, with each qubit pad connected to $X$ couplers. As illustrated in Fig.~\ref{fig: schematic}(a), the conventional two-dimensional grid architecture corresponds to $X=2$. Within the framework of circuit QED, the node fluxes of the two qubit pads are denoted by $\Phi_{a}$ and $\Phi_{b}$, while the node flux of the $i$-th ($i=1,2$) coupler connected to pad $a$ ($b$) is denoted by $\Phi_{ai}$ ($\Phi_{bi}$). The resulting circuit is described by a capacitance network comprising the qubit/coupler capacitance to ground, the capacitance
between qubit pads $C_{ab}$, the qubit–coupler mutual capacitance $C_{qc}$ and several parasitic capacitances. Specifically, $C_{ag}$, $C_{bg}$, and $C_{cg}$ denote the capacitances of qubit pads $a$, $b$, and the couplers to ground, respectively. The parasitic capacitances include $C_{Xqc}$, between a qubit pad and couplers connected to the opposite qubit pad; $C'_{cci}$, between couplers connected to opposite qubit pads; and $C''_{cc}$, between neighboring couplers connected to the same qubit pad.

Following the standard circuit quantization formalism, the kinetic energy of the system Lagrangian is written as
\begin{widetext}
\begin{align}
T & =\frac{C_{ab}}{2}\left(\dot{\Phi}_{a}-\dot{\Phi}_{b}\right)^{2}+\frac{C_{ag}}{2}\left(\dot{\Phi}_{a}\right)^{2}+\frac{C_{bg}}{2}\left(\dot{\Phi}_{b}\right)^{2}+\frac{C_{cg}}{2}\sum_{q=a,b} \sum_{i=1}^2 \left(\dot{\Phi}_{qi}\right)^{2}+\frac{C_{qc}}{2}\sum_{q=a,b} \sum_{i=1}^2 \left(\dot{\Phi}_{q}-\dot{\Phi}_{qi}\right)^{2}\nonumber\\
& +\frac{C_{Xqc}}{2}\sum_{i=1}^2\left(\dot{\Phi}_{a}-\dot{\Phi}_{bi}\right)^{2}+\frac{C_{Xqc}}{2}\sum_{i=1}^2\left(\dot{\Phi}_{b}-\dot{\Phi}_{ai}\right)^{2}\nonumber\\
 & +\frac{C'_{cc1}}{2}\sum_{i=1}^2\left(\dot{\Phi}_{ai}-\dot{\Phi}_{bi}\right)^{2}+\frac{C'_{cc2}}{2}\sum_{\substack{i,j=1\\ j\neq i}}^{2} \left(\dot{\Phi}_{ai}-\dot{\Phi}_{bj}\right)^{2}+\frac{C''_{cc}}{2}\sum_{\ q=a,b}\left(\dot{\Phi}_{q1}-\dot{\Phi}_{q2}\right)^{2}\nonumber\\
 & =\frac{1}{2}\dot{\mathbf{\Phi}}\mathbf{C}\dot{\mathbf{\Phi}}
\end{align}
\end{widetext}
where $\mathbf{\Phi}=(\Phi_{a},\Phi_{b},\Phi_{a_1},\Phi_{a_2},\Phi_{b_1},\Phi_{b_2})$. For analytical tractability, we consider the fully symmetric case, in which the two qubit pads have identical capacitances to ground, $C_{ag}=C_{bg}$, and the four couplers are identical, sharing the same capacitance to ground and the same mutual capacitances to the surrounding circuit elements. Under these assumptions, the capacitance matrix is obtained as Eq.~\eqref{eq:capacitance_matrix}.

Following the standard decomposition into qubit's differential and common modes, we introduce the transformed qubit flux variables $\Phi_{\pm}=\Phi_{a}\pm\Phi_{b}$. The kinetic energy is then rewritten as $T=\dot{\tilde{\Phi}}\mathcal{C}\dot{\tilde{\Phi}}/2$
where $\tilde{\Phi}=M\Phi$, $\mathcal{C}=(M^{T})^{-1}\mathbf{C}M^{-1}$, and the transformation matrix is
\begin{align}
M=\left(\begin{array}{cccccc}
1 & 1 & 0 & 0 & 0 & 0\\
1 & -1 & 0 & 0 & 0 & 0\\
0 & 0 & 1 & 0 & 0 & 0\\
0 & 0 & 0 & 1 & 0 & 0\\
0 & 0 & 0 & 0 & 1 & 0\\
0 & 0 & 0 & 0 & 0 & 1
\end{array}\right).
\end{align}
Performing the Legendre transformation and expressing in terms of the conjugate charge operators,
$Q_{l}=\partial T/\partial\Phi_{l}$ with $l=+,-, a_1,a_2,b_1,b_2$,
yields the capacitive Hamiltonian
\begin{align}
\hat{H} & =\frac{1}{2}\mathbf{\hat{Q}}\mathcal{C}^{-1}\mathbf{\hat{Q}}
\end{align}
where $\mathbf{\hat{Q}}=(\hat{Q}_{+},\hat{Q}_{-},\hat{Q}_{a_1},\hat{Q}_{a_2},\hat{Q}_{b_1},\hat{Q}_{b_2})$ and  the potential energy has been omitted. Since the common mode does not appear in the potential energy, it represents a free degree of freedom and can be eliminated~\cite{Ding2021_CQED}. Retaining only the terms relevant to the qubit–coupler interaction, the reduced Hamiltonian becomes
\begin{align}
\hat{{H}} & =4E_{C}^{q}\hat{n}_{q}^{2}+\sum_{c} \left(4E_{C}^{c}\hat{n}_{c}^{2}+J_{qc}\hat{n}_{q}\hat{n}_{c} \right)
\end{align}
where $\hat{n}_q=\hat{Q}_{-}/2e$ and $\hat{n}_c=\hat{Q}_c/2e$ ($c\in \{a_1,a_2,b_1,b_2\}$) are the Cooper-pair number operators of the qubit and couplers, respectively. The charging energies $E_C^{q}$ and $E_C^{c}$ are determined by the diagonal elements of $\mathcal{C}^{-1}$. Defining the effective qubit capacitance through
$4E_{C}^{q}=2e^{2}/C_{q}^{\mathrm{eff}}$, and $C_{q}^{\mathrm{eff}}=1/(\mathcal{C}^{-1})_{22}$, 
we obtain
\begin{align} \label{eq:Cq_eff}
C_{q}^{\mathrm{eff}} =C_{ab}+ \left[\frac{1}{C^\mathrm{eff}_{ag}} + \frac{1}{C^\mathrm{eff}_{bg}} \right]^{-1} 
\end{align}
with
\begin{equation}
\begin{aligned} \label{eq:Cd_Cbg}
C^\mathrm{eff}_{ag}=C^\mathrm{eff}_{bg} 
=& C_{bg}+\frac{2(C_{cg}+2C_{cc})(C_{qc}+C_{Xqc})}{C_{d}}
\\ &+\frac{8C_{qc}C_{Xqc}}{C_{d}}
\end{aligned}
\end{equation}
where $C_{d}=C_{cg}+2C_{cc}+C_{qc}+C_{Xqc}$, $C_{cc}=C'_{cc1}+C'_{cc2}$. 

Next, the qubit–coupler coupling strength is obtained from the off-diagonal element of the inverse capacitance matrix, $J_{qc}=4e^{2}(\mathcal{C}^{-1})_{23}$, which yields
\begin{align} \label{eq:Jqc}
J_{qc} ={2e^{2}}\frac{C_{qc}-C_{Xqc}}{C_{d}C_{q}^{\mathrm{eff}}}.
\end{align}
Substituting the relations $1/C_d = (C_{q}^{\mathrm{eff}}-C_{ab}-C_{bg}/2)/[(C_{cg}+2C_{cc})(C_{qc}+C_{Xqc})+4C_{qc}C_{Xqc}]$ together with $C_{bg}^\mathrm{eff} =2 (C^\mathrm{eff}_q - C_{ab})$ obtained from Eqs.~\eqref{eq:Cq_eff} and \eqref{eq:Cd_Cbg}, into Eq.~\eqref{eq:Jqc}, we obtain
\begin{align}
J_{qc}
 & ={{2e^{2}}\frac{C_{qc}-C_{Xqc}}{(C_{cg}+2C_{cc})(C_{qc}+C_{Xqc})+4C_{qc}C_{Xqc}}}\nonumber\\&\quad\times\frac{C_{q}^{\mathrm{eff}}-C_{ab}-C_{bg}/2}{C_{q}^{\mathrm{eff}}}\nonumber\\
 & ={\frac{2e^{2}}{C_{cg}}\frac{1-r_{1}}{1+r_{1}+4r_{2}}\times\frac{C_{cg}}{C_{cg}+2C_{cc}}}\nonumber\\&\quad\times\left(1-\frac{C_{ab}}{C_{q}^{\mathrm{eff}}}\right)\times \left(1-\frac{C_{bg}}{C_{bg}^{\mathrm{eff}}}\right)\nonumber\\
 & =\frac{2e^{2}}{C_{cg}}\eta_{qc}\eta_{cc}\left(1-\frac{C_{ab}}{C_{q}^{\mathrm{eff}}}\right)\left(1-\frac{C_{bg}}{C_{bg}^{\mathrm{eff}}}\right)
\end{align}
where $\eta_{qc}=(1-r_{1})/(1+r_{1}+4r_{2}),\ \eta_{cc}=(1+2C_{cc}/C_{cg})^{-1}$ 
with $r_{1}=C_{Xqc}/C_{qc}$, $r_{2}=C_{Xqc}/(C_{cg}+2C_{cc})$. Equivalently, for the symmetric geometry, $J_{qc}$ can be written as
\begin{equation}
    J_{qc} = \frac{2e^{2}}{C_{cg}} \eta_{cc}\eta_{qc} \left(  1-\frac{C_{ab}}{C_{q}^{\mathrm{eff}}}-\frac{C_{bg}/2}{C_{q}^{\mathrm{eff}}} \right),
\end{equation}
which makes explicit that both the inter-pad capacitance $C_{ab}$ and the qubit-to-ground capacitance $C_{bg}$ reduce the achievable coupling through their respective contributions to the effective qubit capacitance.

In the symmetric geometry, the parasitic capacitance $C''_{cc}$ between neighboring couplers connected to the same qubit pad does not contribute to either $C_q^{\mathrm{eff}}$  or $J_{qc}$, whereas the inter-pad parasitic capacitances $C'_{cci}$ do. In the absence of $C'_{cci}$ and $C_{Xqc}$, the expressions for $C_q^\mathrm{eff}$ and $J_{qc}$ reduce to the compact forms summarized in Table~\ref{tab:qc_config}. Generalizing the derivation to arbitrary qubit pad connectivity $X$ preserves the analytical form of the result, with the prefactor replaced by $4e^{2}/(XC_{cg})$. Consequently, the ideal qubit–coupler coupling upper bound exhibits a $1/X$ dependence. The dependence of $J_{qc}$ on the qubit pad connectivity $X$ is identical for the remaining two configurations considered in the following subsections and is therefore omitted for brevity.

\subsection{Grounded qubit with grounded couplers}\label{sec:groundQ_groundC}
We consider a grounded qubit coupled to four grounded couplers (namely $X=4$), as shown in Fig.~\ref{fig: schematic}(b). Applying the circuit QED procedure introduced above, the capacitance matrix is  
\begin{align}
\mathbf{C} & =\left(\begin{array}{ccccc}
C_{q} & -C_{qc} & -C_{qc} & -C_{qc} & -C_{qc}\\
-C_{qc} & C_{c} & -C''_{cc1} & -C''_{cc2} & -C''_{cc1}\\
-C_{qc} & -C''_{cc1} & C_{c} & -C''_{cc1} & -C''_{cc2}\\
-C_{qc} & -C''_{cc2} & -C''_{cc1} & C_{c} & -C''_{cc1}\\
-C_{qc} & -C''_{cc1} & -C''_{cc2} & -C''_{cc1} & C_{c}
\end{array}\right),
\end{align}
where $C_{q}=C_{bg}+4C_{qc}$ , $C_{c}=C_{cg}+C_{qc}+2C''_{cc1}+C''_{cc2}$. Here, $C''_{cc1}$ and $C''_{cc2}$ denote the parasitic capacitances between the adjacent and next-nearest-neighbor couplers, respectively.

The qubit effective capacitance is determined by the diagonal element of the inverse capacitance matrix associated with the qubit mode, $C_{q}^{\mathrm{eff}}=1/(\mathcal{C}^{-1})_{11}$, which yields
\begin{align} \label{eq:Cqeff_gnd}
C_{q}^{\mathrm{eff}} & =C_{bg}+\frac{4C_{cg}C_{qc}}{C_{cg}+C_{qc}}.
\end{align}
The corresponding qubit–-coupler coupling strength is obtained from the off-diagonal element of the inverse capacitance matrix, $J_{qc}=4e^{2}(\mathcal{C}^{-1})_{12}$, which yields
\begin{align} \label{eq:Jqc_gnd}
J_{qc} ={4e^{2}}\frac{C_{qc}}{(C_{cg}+C_{qc})C_{q}^{\mathrm{eff}}}
\end{align}
Using the relation $C_{qc}/(C_{cg}+C_{qc})=(C_q^\mathrm{eff}-C_{bg})/(4C_{cg})$ obtained from Eq.~\eqref{eq:Cqeff_gnd}, we obtain
\begin{align}
J_{qc} =\frac{e^{2}}{C_{cg}} \left(1-\frac{C_{bg}}{C_{q}^{\mathrm{eff}}}\right).
\end{align}

\subsection{Concentric floating qubit with grounded couplers}\label{sec:asyQ_groundC}

We consider a concentric floating qubit coupled to four grounded couplers ($X=4$), where all couplers are connected to a single qubit pad $b$, as shown in Fig.~\ref{fig: schematic}(c). Applying the circuit QED procedure introduced above, the capacitance matrix is
\begin{align}
\mathbf{C} & =\left(\begin{array}{cccccc}
C_{a} & -C_{ab} & 0 & 0 & 0 & 0\\
-C_{ab} & C_{b} & -C_{qc} & -C_{qc} & -C_{qc} & -C_{qc}\\
0 & -C_{qc} & C_{c} & -C''_{cc1} & -C''_{cc2} & -C''_{cc1}\\
0 & -C_{qc} & -C''_{cc1} & C_{c} & -C''_{cc1} & -C''_{cc2}\\
0 & -C_{qc} & -C''_{cc2} & -C''_{cc1} & C_{c} & -C''_{cc1}\\
0 & -C_{qc} & -C''_{cc1} & -C''_{cc2} & -C''_{cc1} & C_{c}
\end{array}\right),
\end{align}
where $C_{a}=C_{ag}+C_{ab}$, $C_{b}=C_{bg}+C_{ab}+4C_{qc}$, and
$C_{c}=C_{cg}+C_{qc}+2C''_{cc1}+C''_{cc2}$. The definitions of $C''_{cc1}$ and $C''_{cc2}$ are the same as those in the previous subsection. For this reduced analytical model, we neglect the capacitances between the qubit pad $a$ and coupler pads connected to pad $b$. 

The effective qubit capacitance is determined from the diagonal element of the inverse capacitance matrix associated with the qubit mode, $C_{q}^{\mathrm{eff}}=1/(\mathcal{C}^{-1})_{22}$, yielding
\begin{align} \label{eq:Cqeff_concentric}
C_{q}^{\mathrm{eff}} =C_{ab}+ \left[\frac{1}{C^\mathrm{eff}_{ag}} + \frac{1}{C^\mathrm{eff}_{bg}} \right]^{-1} 
\end{align}
with 
\begin{align} \label{eq:Cbgeff_concentric}
C_{ag}^{\mathrm{eff}} & =C_{ag},\nonumber\\
C_{bg}^{\mathrm{eff}} & =C_{bg}+4\frac{C_{cg}C_{qc}}{C_{cg}+C_{qc}}.
\end{align}
The corresponding qubit–coupler coupling strength is obtained from the off-diagonal element of the inverse capacitance matrix, $J_{qc}=4e^{2}(\mathcal{C}^{-1})_{23}$, which yields
\begin{align}
J_{qc} & ={4e^{2}}\frac{C_{qc}}{C_{d}C_{q}^{\mathrm{eff}}},
\end{align}
where $C_{d}=(C_{cg}+C_{qc})(1+C_{bg}^\mathrm{eff}/C_{ag}^\mathrm{eff})$. Using the relations $(C_{cg}+C_{qc})=(4C_{cg}C_{qc})/(C_{bg}^\mathrm{eff}-C_{bg})$, $(1+C_{bg}^\mathrm{eff}/C_{ag}^\mathrm{eff})=C_{bg}^\mathrm{eff}/(C_{q}^\mathrm{eff}-C_{ab})$ obtained from the above Eqs.~\eqref{eq:Cqeff_concentric} and \eqref{eq:Cbgeff_concentric}, the coupling strength can be recast into the same form solved for the previous two configurations,
\begin{align}
J_{qc} & ={4e^{2}} \frac{C_{bg}^\mathrm{eff}-C_{bg}}{4C_{cg}C_{qc}}  \times \frac{C_{q}^\mathrm{eff}-C_{ab}}{C_{bg}^\mathrm{eff}} \times \frac{C_{qc}}{C_{q}^{\mathrm{eff}}}\nonumber\\
 & =\frac{e^{2}}{C_{cg}}\left(1-\frac{C_{ab}}{C_{q}^{\mathrm{eff}}}\right)\left(1-\frac{C_{bg}}{C_{bg}^{\mathrm{eff}}}\right).
\end{align}

\section{Electromagnetic Simulations: Setup, Results and Discussion}

In this section, we present the technical details of electromagnetic (EM) simulation employed throughout this work. 
Sec.~\ref{sec: EM_general_setup} introduces the EM simulation setup and workflow for extracting the  capacitance matrix from the design layouts, as well as the subsequent circuit-quantization procedure. Sec.~\ref{sec: EM_1q4c} validates our main analytical result, i.e., Eq.\eqref{eq: analytical_Jqc}, using a minimal unit comprising one floating qubit and four grounded couplers. Sec.~\ref{sec:r_bg} systematically investigates the geometric dependence of the ratio $C_{bg}/C_{bg}^\mathrm{eff}$. Sec.~\ref{subsec:comparison} presents the detailed capacitances and capacitive participation ratios of optimized and conventional 5q4c layout designs.

\subsection{EM Simulation Setup}\label{sec: EM_general_setup}

This section describes the three-dimensional EM model and simulation configuration adopted throughout this work, unless otherwise specified. All EM simulations are performed in the electrostatic limit to extract the Maxwell capacitance matrix $\mathbf{C}_\mathrm{M}$ on 2D planar quantum chip layouts comprising four components enclosed by a vacuum box: the qubit pad(s), coupler pad(s), surrounding ground plane, and substrate. The superconducting metal layer is modeled as a zero-thickness perfect electric conductor (PEC) and placed on a square sapphire substrate of width $6\,\mathrm{mm}$ and thickness $430\,\mu\mathrm{m}$ with a relative permittivity of 10.2. For each qubit or coupler, an etched opening in the metal layer electrically isolates the central qubit and coupler pads from the surrounding ground plane. The common simulation parameters are summarized in Table~\ref{tab:em_setup}. 

\begin{table}[h]
\caption{Common Parameters of EM Simulation Setups.}
\label{tab:em_setup}
\centering
\begin{tabular}{lll}
\hline\hline
Parameter & Value & Definition \\
\hline

$\varepsilon_r^{\mathrm{sapphire}}$
& 10.2
& Relative permittivity of sapphire \\

$t_\text{sub}$
& $430\,\mu\mathrm{m}$
& Thickness of sapphire substrate \\

$w_\text{sub}$
& $6\,\mathrm{mm}$
& Width of square substrate \\

$w_{\mathrm{gnd}}$
& $6\,\mathrm{mm}$
& Width of square ground etch \\

\hline\hline
\end{tabular}
\\
\end{table}

\begin{figure*}[!t]
    \centering
    \includegraphics[width=0.87\linewidth]{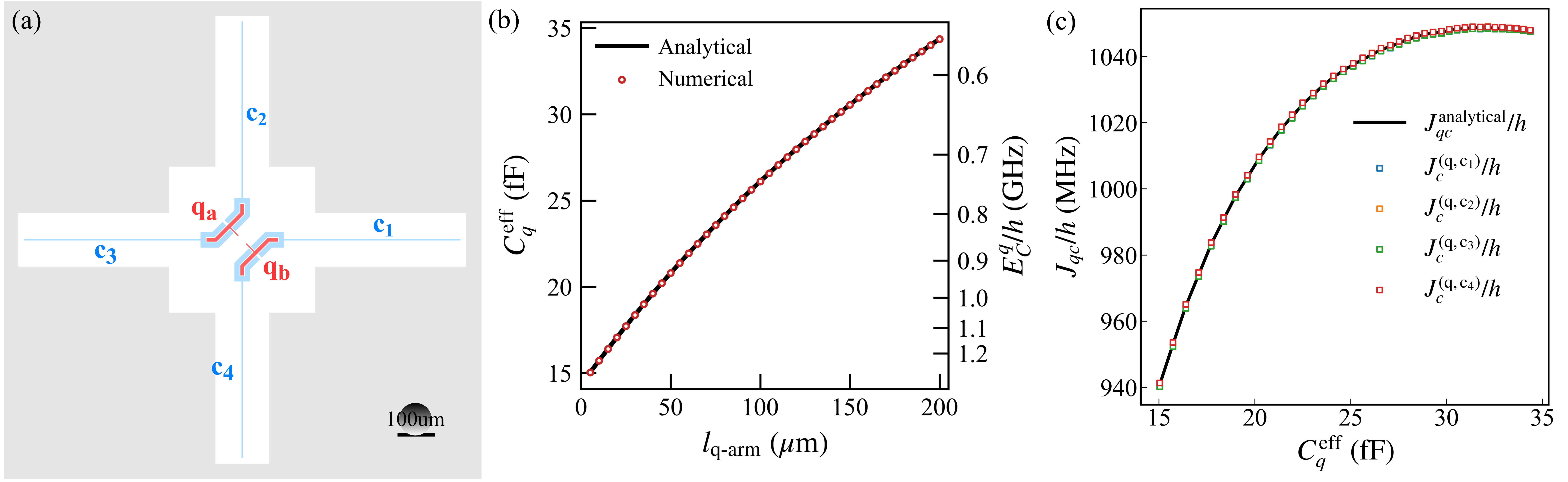}
    \caption{(a) Schematic layout of the device consisting of one floating qubit (red pads, $q_a$ and $q_b$) symmetrically coupled to four grounded couplers (blue pads, $c_1, c_2, c_3,$ and $c_4$). The grounded planes are shown in gray, and the etched regions are white.  (b) Qubit effective capacitance $C_q^{\mathrm{eff}}$ (left axis) and qubit charging energy $E_C^q/h$ (right axis) as a function of the qubit arm length $l_{q\text{-arm}}$, plotted on dual axes to facilitate direct correlation between the geometric capacitance $C_q^{\mathrm{eff}}$ and the Hamiltonian parameter $E_C^q$. The analytical profiles (black solid lines) show good agreement with the numerical results (red open circles).  (c) Qubit--coupler capacitive coupling energy $J_{qc}$ (absolute values) versus $C_q^{\mathrm{eff}}$. Numerical results (colored open squares) match well with the analytical model (black solid line). }
    \label{fig: EM_anlaytical_model}
\end{figure*}

To ensure simulation accuracy, a refined mesh with a maximum element size constraint of $1\,\mu\mathrm{m}$ was employed in the qubit and coupler regions to resolve capacitance contributions. Adaptive refinement was terminated when both the relative electrostatic energy error and the energy variation between consecutive iterations were reduced below $0.5\%$, providing sufficient numerical accuracy for the extraction of the capacitance parameters.

To extract the circuit parameters, we perform circuit quantization using the Maxwell capacitance matrix $\mathbf{C}_{\text{M}}$ obtained directly from EM simulations. The total capacitance matrix is constructed by incorporating the Josephson junction capacitances $\mathbf{C}_{\text{JJ}}$ and additional lumped capacitances $\mathbf{C}_{\text{add}}$ (e.g., those associated with Josephson junction arrays),
\begin{equation}
\mathbf{C}=\mathbf{C}_{\mathrm{M}}+\mathbf{C}_{\mathrm{JJ}}+\mathbf{C}_{\mathrm{add}}.
\end{equation}
Each lumped capacitance connected between nodes $i$ and $j$ is incorporated into $\mathbf{C}$ according to the standard graph-Laplacian construction. Accordingly, the total branch capacitance is
\begin{equation}
C_{ij}=C_{ij}^{\mathrm{geo}}+C_{ij}^{\mathrm{JJ}}+C_{ij}^{\mathrm{add}},
\end{equation}
where $C_{ij}^{\mathrm{geo}}$ denotes the geometric capacitance extracted from $\mathbf{C}_{\mathrm{M}}$, $C_{ij}^{\mathrm{JJ}}$ is the Josephson junction capacitance, and $C_{ij}^{\mathrm{add}}$ represents additional lumped contributions, such as those from Josephson junction arrays. 

The resulting capacitance matrix $\mathbf{C}$ is then analyzed using a spanning-tree graph formalism to extract the effective qubit and coupler capacitances, qubit–coupler coupling strengths, and other circuit parameters. This procedure yields the inverse capacitance matrix $\mathcal{C}^{-1}$, from which the charging energy of node $i$ is given by
\begin{equation}
E_C^{i} = \frac{e^2}{2} \mathcal{C}^{-1}_{ii},
\end{equation}
where the Hamiltonian is parameterized by the charging term $E_C^{i}\hat{n}_i^2$. The off-diagonal elements of $\mathcal{C}^{-1}$  determine the capacitive coupling strength between nodes $i$ and $j$,
\begin{equation}
J_c^{(i,j)} = 4e^2\mathcal{C}^{-1}_{ij},~ (i\neq j).
\end{equation}
This circuit-quantization workflow is used throughout this work to obtain the numerical results of the charging energies and qubit--coupler coupling strengths.

\subsection{Analytical result validation} \label{sec: EM_1q4c}

\begin{table*}[t]
\caption{Design parameters of the layout shown in Fig.~\ref{fig: EM_anlaytical_model}(a)}
\label{tab:em_analytical_validation}
\centering
\begin{tabular}{llll}
\hline\hline
Component & Parameter & Value & Definition \\
\hline

\multirow{6}{*}{Qubit}
& $w_q$
& $10\,\mu\mathrm{m}$
& Qubit connection and arm width \\
& $w_{q\text{-gap}}$
& $90\,\mu\mathrm{m}$
& Gap width between qubit pads \\
& $l_{q\text{-conn}}$
& $w_q+w_{q\text{-gap}}$
& Qubit connection section length \\
& $l_{q\text{-arm}}$
& $5\text{--}200\,\mu\mathrm{m},\ \Delta=5\,\mu\mathrm{m}$
& Qubit arm length \\
& $w_{q\text{-hook}}$
& $3\,\mu\mathrm{m}$
& Qubit hook width \\
& $w_{q\text{-hook-gap}}$
& $3\,\mu\mathrm{m}$
& Gap width between qubit hook tips \\

\hline

\multirow{3}{*}{Coupler}
& $w_{qc\text{-gap}}$
& $2\,\mu\mathrm{m}$
& Qubit--coupler gap width \\
& $w_{c\text{-couple}}$
& $14\,\mu\mathrm{m}$
& Coupler coupling section width \\
& $w_{c\text{-ext}}$
& $4\,\mu\mathrm{m}$
& Coupler extension section width \\

\hline

\multirow{5}{*}{Ground}
& $w_{q\text{-gnd-padding}}$
& $100\,\mu\mathrm{m}$
& Padding from and orthogonal to qubit pad tips \\
& $w_{q\text{-gnd-etch}}$
& $\sqrt{2}\,l_{q\text{-conn}}
+2\left(l_{q\text{-arm}}
+w_{q\text{-gnd-padding}}\right)$
& Width of square etched region surrounding qubit \\
& $w_{c\text{-gnd-padding}}$
& $70\,\mu\mathrm{m}$
& Orthogonal padding from coupler pad \\
& $w_{c\text{-gnd-etch}}$
& $w_{c\text{-ext}}
+2\,w_{c\text{-gnd-padding}}$
& Width of rectangular etched region surrounding coupler \\

\hline\hline
\end{tabular}
\end{table*}

\begin{figure*}[t]
    \centering
    \includegraphics[width=0.91\linewidth]{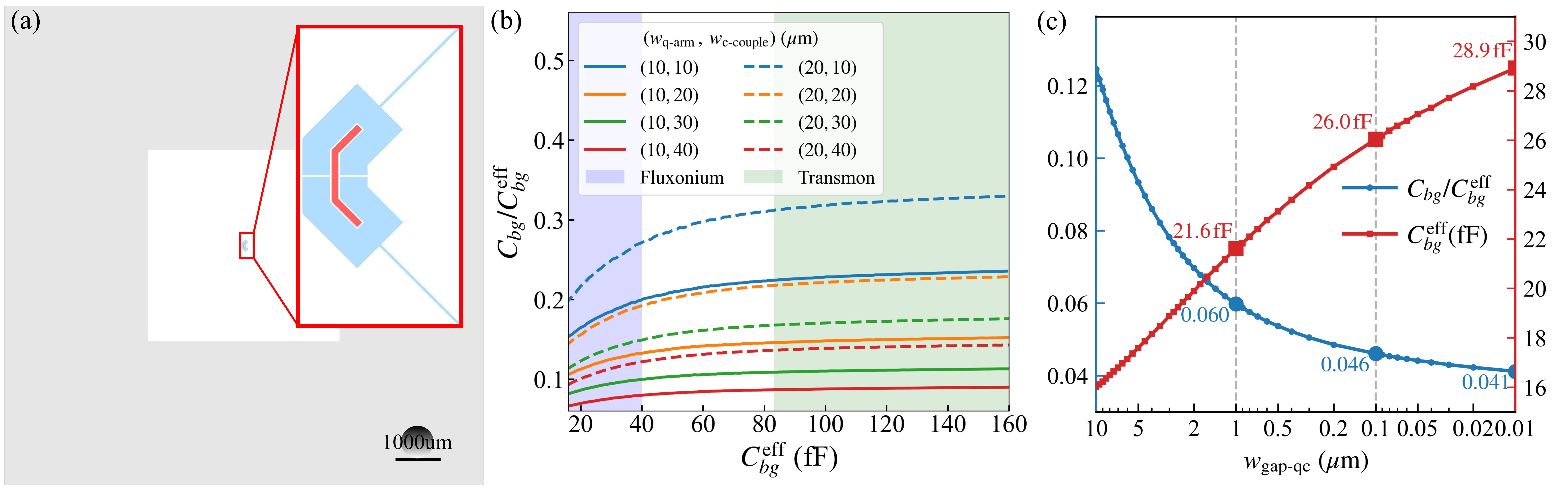}
    \caption{
    (a) Representative $1\mathrm{p}2\mathrm{c}$ configuration.
    (b) $C_{bg}/C_{bg}^\mathrm{eff}$ as a function of $C_{bg}^\mathrm{eff}$
    for varying $w_{q\text{-arm}}$ and $w_{c\text{-couple}}$.
    (c) $C_{bg}/C_{bg}^\mathrm{eff}$ and $C_{bg}^\mathrm{eff}$ as functions
    of the qubit--coupler gap width $w_{\text{gap-}qc}$.
    Representative values at selected gap widths are labeled.}
    \label{fig:em_eng_gapqc_sweep}
\end{figure*}

The procedure presented in the previous section provides a direct numerical realization of the capacitance-participation framework developed in the main text, allowing the analytical predictions to be evaluated using the full extracted
capacitance network. Based on the general EM simulation setup, we construct a minimal model consisting of 1 dumbbell floating qubit and 4 grounded couplers (1q4c) to validate the analytical equation derived, i.e., Eq.~\eqref{eq: analytical_Jqc}. As shown in Fig.~\ref{fig: EM_anlaytical_model}(a), two symmetrical qubit pads with hooks are positioned at the center of the layout, with a square-etched region. Each pad is coupled by two identical coupler pads with rectangular ecthed regions. This symmetrical configuration aligns with the capacitance matrix, namely Eq.~\eqref{eq:capacitance_matrix}. The coupler extension segments are orthogonal to each other, which supports a two-dimensional grid architecture. The geometric design variables and the corresponding values used in the EM simulations are summarized in Table~\ref{tab:em_analytical_validation}.
To validate the analytical theory, we benchmark it against circuit QED calculations. Both approaches use the same Maxwell capacitance matrix extracted from electrostatic electromagnetic simulations and differ only in the subsequent parameter extraction procedure. In the analytical approach, the relevant geometric capacitances are first corrected by explicitly incorporating the capacitances of the Josephson junctions and Josephson junction arrays, e.g., $C_{ab}=C_{ab}^{\rm geo}+C_{ab}^{\rm JJ}+C_{ab}^{\rm JJA}$ and $C_{cg}=C_{cg}^{\rm geo}+C_{cg}^{\rm JJ}$, and are then substituted into the analytical expressions derived in Sec.~\ref{sec: symm_fqgc}. By contrast, the numerical approach applies the full network-reduction and circuit-quantization procedure described in Sec.~\ref{sec: EM_general_setup} directly to the complete capacitance matrix. As shown in Figs.~\ref{fig: EM_anlaytical_model}(b) and \ref{fig: EM_anlaytical_model}(c), the analytical predictions (solid lines) are in excellent agreement with the numerical results (open symbols) over the entire parameter range, validating the analytical result against full-network circuit quantization using the same extracted capacitance matrix.

\subsection{Electromagnetic model and geometric dependence of $C_{bg}/C_{bg}^\mathrm{eff}$} \label{sec:r_bg}

First of all, we introduce the technical details concerning the EM model used in Fig.~\ref{fig: em-eng}(b). For different qubit connectivity, the qubit pads are designed to have 1, 2, and 3 arms in each setup with equal width  $w_q$. The coupler pads are arranged to closely surround the qubit pad in order to maximize the qubit-to-coupler capacitance while suppressing the qubit-to-ground capacitance. This is mainly achieved by increasing coupler coupling section width,  $w_{c\text{-couple}}$, given certain qubit--coupler coupling length. The qubit--coupler pad separation, denoted as  $w_{qc\text{-gap}}$, is set to  $2\,\mu \rm m$, representing a practical lower bound determined by fabrication constraints, unless otherwise specified. In multi-coupler configurations, the qubit includes connection sections that bridge adjacent arms. To ensure complete enclosure of the qubit pad, the coupling sections of the couplers are extended beyond the qubit arm length. The gaps between neighboring couplers are chosen to be equal to the qubit--coupler gap width. Apart from the coupling section, the coupler includes an extension segment with a width of $w_c$ and a length of $l_c$. This extension is introduced to emulate the geometry of a realistic coupler and to account for its associated coupler-to-ground capacitance contribution. To ensure adequate ground-metal coverage around the $4~\mathrm{mm}\times4~\mathrm{mm}$ etched region, the substrate and ground dimensions are both set to $w_{\mathrm{sub}}=w_{\mathrm{gnd}}=10~\mathrm{mm}$.

Next, we investigate the geometric dependence of $C_{bg}/C_{bg}^\mathrm{eff}$. Beyond its dependence on the qubit capacitance budget and qubit connectivity as shown in Sec.~\ref{sec:analytical_derivation}, we further examine its dependence on the geometric design parameters. We adopt the EM simulation setup described above and use the $1\mathrm{p}2\mathrm{c}$ configuration as a representative model, shown in Fig.~\ref{fig:em_eng_gapqc_sweep}(a).
The capacitance matrices are then extracted to compute $C_{bg}/C_{bg}^\mathrm{eff}$, which characterize the electrostatic screening provided by the surrounding coupler pads. Physically, a smaller ratio indicates that a larger fraction of the qubit-to-ground capacitance is established through indirect coupling paths mediated by the couplers, rather than through the direct qubit-to-ground capacitance. From the relation
\begin{equation}
C_{bg}^{\mathrm{eff}}
=
C_{bg}+
\sum_{i=1}^{X}
\left(
C_{bc,i}^{-1}+
C_{cg,i}^{-1}
\right)^{-1},
\label{eq:Cbg_eff}
\end{equation}
it follows that $C_{bg}/C_{bg}^{\mathrm{eff}}$ decreases as the direct capacitance $C_{bg}$ decreases or as the indirect coupling path, characterized by $C_{bc}$ and $C_{cg}$, becomes stronger. Among geometric design parameters, the electrostatic screening effect is mainly governed by three key variables: the width of the qubit coupling arm $w_{q\text{-arm}}$, the width of the coupler coupling section $w_{c\text{-couple}}$, and the qubit--coupler gap $w_{qc\text{-gap}}$. Then, we study how the ratio $C_{bg}/C_{bg}^{\mathrm{eff}}$ depends on these dominant geometric parameters.

To elucidate the joint influence of coupling section widths on screening performance, we varied parameter pair ($w_{q\text{-arm}}$, $w_{c\text{-couple}}$) while keeping the ground etch size, qubit/coupler pad widths, qubit connection segment length, and the qubit--coupler gap fixed. As illustrated in Fig.~\ref{fig:em_eng_gapqc_sweep}(b), the length of the qubit arm, $l_{q\text{-arm}}$, is varied continuously to simultaneously modulate both the qubit shunt capacitance and the qubit-to-coupler capacitance. Specifically, we evaluate two discrete values for the qubit arm width, $w_{q\text{-arm}} = 10\,\mu\mathrm{m}$ and $20\,\mu\mathrm{m}$, represented by solid and dashed lines, respectively. For each arm width, the coupler coupling section width $w_{c\text{-couple}}$ spans across $10\,\mu\mathrm{m}$ (blue), $20\,\mu\mathrm{m}$ (orange), $30\,\mu\mathrm{m}$ (green), and $40\,\mu\mathrm{m}$ (red), where the red solid line corresponds to the configuration discussed in Sec.\ref{sec:MS_DP}. Two prominent trends emerge from the simulated trajectories: under a fixed target $C_{bg}^{\mathrm{eff}}$, widening the coupler coupling section ($w_{c\text{-couple}}$) or narrowing the qubit arm ($w_{q\text{-arm}}$) systematically suppresses the $C_{bg}/C_{bg}^{\mathrm{eff}}$ ratio. Consequently, a geometric configuration featuring a wider coupler section paired with a narrower qubit arm minimizes this ratio, thereby enhancing the electrostatic screening effect and reducing cross-coupling to the environment.

As illustrated in Fig.~\ref{fig:em_eng_gapqc_sweep}(c), we further investigate the impact of the qubit--coupler gap width ($w_{\text{gap}\mbox{-}{qc}}$) by tracking both the capacitance ratio $C_{bg}/C_{bg}^{\mathrm{eff}}$ and the absolute effective capacitance $C_{bg}^{\mathrm{eff}}$, while fixing the qubit arm length at $l_{q\text{-arm}} = 50\,\mu\mathrm{m}$. As the gap width decreases, the ratio $C_{bg}/C_{bg}^{\mathrm{eff}}$ exhibits a decelerating decline on a logarithmic scale. Physically, this suppression indicates that as the coupler is brought into extreme proximity with the qubit, the enhanced electrostatic screening effect begins to saturate due to the geometric constraints of the local field distribution. To establish a fair comparison with the results presented in Sec.~\ref{sec:MS_DP}, we mark three representative gap regimes via vertical dashed lines ($1\,\mu\mathrm{m}$, $100\,\mathrm{n}\mathrm{m}$, and $10\,\mathrm{n}\mathrm{m}$). These configurations yield corresponding benchmarks of $[w_{{qc\text{-gap}}}, C_{bg}/C_{bg}^{\mathrm{eff}}, C_{bg}^{\mathrm{eff}}] = [1\,\mu\mathrm{m}, 0.060, 21.6\,\mathrm{fF}]$, $[0.1\,\mu\mathrm{m}, 0.046, 26.0\,\mathrm{fF}]$, and $[0.01\,\mu\mathrm{m}, 0.041, 28.9\,\mathrm{fF}]$. Crucially, maintaining a target $C_{bg}^{\mathrm{eff}}$ is required to fix the qubit charging energy $E_C$ as a key design specification. Under identical $C_{bg}^{\mathrm{eff}}$ targets of $21.6\,\mathrm{fF}$, $26.0\,\mathrm{fF}$, and $28.9\,\mathrm{fF}$, the main-text approach (modulating $l_{q\text{-arm}}$) results in higher ratios of $0.071$, $0.073$, and $0.075$, respectively. This comparison demonstrates that scaling down the qubit--coupler gap provides an effective geometric degree of freedom to minimize the $C_{bg}/C_{bg}^{\mathrm{eff}}$ ratio for identical target qubit parameters. In practice, however, utilizing this parameter space is strictly fabrication-limited, as achieving submicrometer gaps pushes the boundaries of standard lithography and etching resolution.

\begin{widetext}
\subsection{Decomposition of capacitance budget consumption in layout designs}\label{subsec:comparison}

Table~\ref{tab:comparison} details the physical parameters, extracted capacitance components, and participation ratios corresponding to each attenuation stage of the capacitive budget decomposition in Fig.~\ref{fig: application_5q4c}(d). By comparing the conventional and optimized layouts under identical target charging energies, it illustrates how combined geometric wrapping and junction process optimization systematically mitigate parasitic loading in both the inter-pad ratio ($C_{ab}/C_q^\mathrm{eff}$) and the pad-to-ground ratio ($C_{bg}/C_{bg}^\text{eff}$). These structural improvements ultimately yield the substantial enhancement in achievable qubit–coupler coupling strength $J_{qc}$.

\begin{table*}[htbp]
\caption{Comparison between the optimized and conventional designs.}
\label{tab:comparison}
\centering
Qubit and coupler charging energies: $E_C^q/h=1.1\,\text{GHz}$, $E_C^c/h=E_C^{c_a}/h=E_C^{c_b}/h=0.3\,\text{GHz}$
\begin{tabular}{clcc}
\hline
\hline
Category & Term & Conventional Design & Optimized Design \\
\hline

\multirow{2}{*}{Junction}
& $hC^\text{JJ}/E_J$ (fF/GHz)
& 0.8
& 0.6 \\
& $C^\text{JJA}$ (fF)
& 2.5
& 1 \\
\hline

\multirow{3}{*}{Coupler-to-ground}
& $C_{cg}^\text{geo}$ (fF)
& 40.71
& 37.87 \\
& $C_{cg}^\text{JJ}$ (fF)
& 12
& 9 \\
& \textbf{${4e^{2}}/{(XhC_{cg})}$} (MHz)
& {1470}
& \textbf{1653} \\

\hline

\multirow{4}{*}{Inter-pad}
& $C_{ab}^{\rm geo}$ (fF)
& 0.841
& 0.189\\
& $C_{ab}^{\rm J,\Sigma}$ (fF)
& 6.22
& 3.79 \\
& $C_q^\mathrm{eff}$ (fF)
& 17.61
& 17.61 \\
& $C_{ab}/C_q^\mathrm{eff}$
& 0.401
& \textbf{0.226} \\
\hline

\multirow{3}{*}{Qubit-to-ground}
& $C_{bg}$ (fF)
& 6.628
& 2.14\\
& $C_{bg}^\mathrm{eff}$ (fF)
& 21.1 
& 27.26\\
& $C_{bg}/C_{bg}^\mathrm{eff}$
& 0.314 
& \textbf{0.0785} \\

\hline

\multirow{2}{*}{Qubit-to-coupler}
& $C_{qc}$ (fF)
& 7.52 
& 15.07\\ 
& $J_{qc}/h$ (MHz)
& 472.1
& \textbf{899.0}\\

\hline
\hline
\end{tabular}
\end{table*}
\end{widetext}

\section{Numerical Simulation of MAP Gate Fidelity}
\label{sec:approach_MAPinFIDELITY}
This section provides the theoretical framework for evaluating the MAP gate fidelity of the optimized 5q4c design shown in Sec.~\ref{sec:5Q4C} and the 7q6c design discussed in Sec.~\ref{sec:7Q6C}. The derivation presented below closely follows Ref.~\cite{Chan2026_F_DTC}, with the essential procedures summarized here for completeness. In particular, Sec.~\ref{subsec:System_Hamiltonian} constructs the system Hamiltonian, incorporating the fluxonium qubits, double-transmon couplers (DTCs), their nearest-neighbor interactions, and the external microwave drive. Sec.~\ref{subsec:Leakage_Analysis} applies time-dependent perturbation theory to derive analytical expressions for the leakage probabilities. Sec.~\ref{subsec:Gate_Fidelity} derives an analytical estimate of the average gate infidelity by constructing an effective unitary in an expanded Hilbert space and modeling coherent leakage as a single Kraus operator.

\subsection{System Hamiltonian}\label{subsec:System_Hamiltonian}

The scalable quantum processor comprises fluxonium qubits coupled via DTCs. As the fundamental building block, we consider the minimal subsystem consisting of two fluxonium qubits connected by a single DTC. The corresponding Hamiltonian is given by
\begin{equation}
    \hat{H}_{\text{total}} = \hat{H}_{\text{sys}} + \hat{H}_d,
\end{equation}
where $\hat{H}_{\text{sys}}$  represents the static Hamiltonian consisting of the fluxoniums, the DTCs, and the capacitive couplings between them, and  $\hat{H}_d$  denotes the Hamiltonian for the time-dependent external microwave control. Furthermore, the system Hamiltonian, $\hat{H}_\text{sys}$, can be decomposed into bare component energies and their mutual interactions:

\begin{equation}
    \begin{aligned}
        \hat{H}_{\text{sys}} &= \sum_{j=1}^2 \hat{H}_{q_j}(E_{C}^{q_j}, E_{J}^{q_j},E_{L}^{q_j})
        \\
        &+  \hat{H}_{c}(E_{C}^{c_{a}}, E_{C}^{c_{b}}, E_{J}^{c_{a}}, E_{J}^{c_{b}}, E_{J}^{c_{ab}}, J_c^{(c_{a},\,c_{b})})
        \\
        &+ \sum_{j=1}^2 \hat{H}_{q_j,c}^{\text{int}}(J_{c}^{(q_j,c)}).
    \end{aligned}.
\end{equation}
In this notation,  $\hat{H}_{q_j}$  corresponds to the $j$-th fluxonium qubit,  $\hat{H}_{c}$ to the DTC, and  $\hat{H}_{q_j,c}^\text{int}$  to the interaction between them. The bare Hamiltonian for the $j$-th fluxonium qubit is parameterized by its charging energy  $E_{C}^{q_j}$, Josephson energy  $E_{J}^{q_j}$, and inductive energy  $E_{L}^{q_j}$, yielding
\begin{equation}
\hat{H}_{q_j} = 4E_{C}^{q_j}\hat{n}_{q_j}^2 - E_{J}^{q_j}\cos\hat{\phi}_{q_j} + \frac{E_{L}^{q_j}}{2}\left(\hat{\phi}_{q_j} - \phi_{\text{ext},{q_j}}\right)^2.
\end{equation}
In this expression,  $\hat{n}_{q_j}$  and  $\hat{\phi}_{q_j}$  denote the conjugate charge and phase operators, respectively, and  $\phi_{\text{ext},q_j}$  represents the external magnetic flux threading the qubit circuit. For the DTC, the Hamiltonian  $\hat{H}_c$  incorporates the dynamics of its two constituent transmons $c_m$ (indexed by $m$ = a, b) and the coupling between them:
\begin{equation}
    \begin{aligned}
        \hat{H}_{c} = &\sum_{m=a,b} \left[ 4E_{C}^{c_m} \hat{n}_{c_m}^{2} - E_{J}^{c_m} \cos \left( \hat{\phi}_{c_m} \right) \right] \\
        &- E_{J}^{c_{ab}} \cos \left( \hat{\phi}_{c_b} - \hat{\phi}_{c_a} - \phi_{\text{ext},{c}} \right) \\
        &+ J_{c}^{(c_a,\,c_b)} \hat{n}_{c_a} \hat{n}_{c_b}.
    \end{aligned}
\end{equation}
Here,  $E_{C}^{c_m}$  and  $E_{J}^{c_m}$  are the respective charging and Josephson energies of the $m$-th transmon mode, while  $E_{J}^{c_{ab}}$  and  $J_{c}^{(c_a,\,c_b)}$  quantify the inter-transmon Josephson energy and capacitive coupling energy. The operators  $\hat{n}_{c_m}$  and  $\hat{\phi}_{c_m}$  correspond to the charge and phase of the individual transmon modes, and  $\phi_{\text{ext},c}$  is the external flux bias applied to the DTC loop. The capacitive coupling between the $j$-th fluxonium and its neighboring DTC is governed by
\begin{equation}
    \hat{H}_{q_j,c}^{\text{int}} = J_{c}^{(q_j,\,c)}\hat{n}_{q_j}\hat{n}_{c_m}.
\end{equation}
Here,  $J_{c}^{(q_j,\,c)}$  dictates the strength of the mutual capacitive interaction between the charge operator  $\hat{n}_{q_j}$  of the fluxonium and the charge operator  $\hat{n}_{c_m}$  of the $m$-th transmon mode within the DTC. For theoretical clarity, our analysis assumes strict nearest-neighbor interactions (e.g. $q_1$ and $c_a$, $c_a$ and $c_b$, $c_b$ and $q_2$ given the $q_1\text{-}c_a\text{-}{c_b}\text{-}q_2$ layout arrangement), thereby omitting any stray coupling between non-adjacent elements.

The external control that applies microwave pulses to the fluxoniums to execute single- and two-qubit gates is formulated as
\begin{equation}
    \hat{H}_d(t) = \Omega_{d}^{q_j}(t) \left( \hat{n}_{q_j} + \gamma_{q_j,c} \hat{n}_{c_m} \right).
\end{equation}
In this model,  $\Omega_{d}^{q_j}(t)$  represents the time-resolved microwave envelope acting on the $j$-th fluxonium. The term scaled by  $\gamma_{q_j,c}$  captures the unavoidable microwave crosstalk experienced by the adjacent coupler. Specifically,  $\gamma_{q_j,c}$ defines the ratio of the mutual capacitance of the drive to the DTC relative to the qubit.

Defining the system Hamiltonian $\hat{H}_{\text{sys}}$ and the drive term $\hat{H}_d$ establishes the foundation for the subsequent section's analysis of leakage for microwave-activated phase gates in a fluxonium-based multi-qubit system.
For a given parameter set, diagonalizing $\hat{H}_{\text{sys}}$ yields the eigenfrequencies $\omega_\alpha = E_\alpha/\hbar$ corresponding to the eigenstates $\vert{}\alpha\rangle$.

\subsection{Leakage Analysis}
\label{subsec:Leakage_Analysis}

Here, we provide analytical estimates of the leakage probabilities during the two-qubit gate operation. When the coupler frequency is tuned to be near resonance to the plasmon frequency of half-flux fluxonium qubits, the composite system of two qubits coupled by the coupler yields eight distinct transitions between the qubit computational states and the high-energy states that could be selectively chosen to be driven to accumulate a conditional phase. Such a scheme is termed the microwave activated phase (MAP) gate. Achieving high fidelity gate requires selecting the transition that minimizes leakage. We quantify this error using time-dependent perturbation theory, under the assumption that the overall leakage probability remains small. Suppose that the system is subjected to a microwave drive with a cosine envelope. Denoting the transition frequency between two states  $|\alpha\rangle$  and  $|\beta\rangle$  as  $\omega_{\alpha,\beta} = \omega_{\alpha} - \omega_{\beta}$, when the drive frequency  $\omega_d$  is tuned to resonance with the desired transition  $|j\rangle \rightarrow |\beta\rangle$  (such that  $\omega_d = \omega_{j,\beta}$ ), parasitic leakage can occur if another transition  $|k\rangle \rightarrow |\alpha\rangle$  possesses a frequency $\omega_{k,\alpha}$ close to the drive frequency. At the end of the gate pulse, the resulting leakage probability, denoted  $\eta_{j\beta,k\alpha}$, is expressed as:

\begin{equation}
\eta_{j\beta,k\alpha} \leq \left| \frac{1}{t_{\rm gate}} \frac{n_{k\alpha}}{n_{j\beta}} \frac{2\pi}{\Delta_{j\beta,k\alpha}} \frac{1}{1 - [t_{\rm gate} \Delta_{j\beta,k\alpha}/(2\pi)]^2} \right|^2.    
\end{equation}

In this expression,  $t_{\rm gate}$  represents the gate duration, while the detuning between the target and parasitic transitions is given by  $\Delta_{j\beta,k\alpha} = |\omega_{\alpha} - \omega_{k}| - |\omega_{\beta} - \omega_{j}|$ ;  $n_{j\beta} = \langle j|\hat{n}_\mathrm{eff}|\beta\rangle$  and  $n_{k\alpha} = \langle k|\hat{n}_\mathrm{eff}|\alpha\rangle$  correspond to the charge matrix elements of the relevant transitions with  $\hat{n}_\mathrm{eff} = \hat{n}_f - \gamma_{j,c} \hat{n}_c$, where  $\hat{n}_f$  and  $\hat{n}_c$  are the charge operators of a fluxonium and a coupler, respectively. For our calculations, we adopt a crosstalk coefficient of $\gamma_{j,c} = 1.29$, a representative value extracted from the practical layout.

When calculating the leakage-induced gate infidelity in the subsequent section, we assume leakage is entirely driven by the most dominant parasitic path, neglecting sub-dominant channels. To maximize gate fidelity, we select the target MAP transition that inherently minimizes this worst-case leakage. Thus, we define $\eta$ as the dominant leakage probability for the optimal MAP transition:
\begin{equation}
\eta=\min_{j,\beta}\left(\max_{k\neq j,\alpha}\eta_{j\beta, k\alpha}\right).    
\end{equation}

\subsection{Gate Infidelity}
\label{subsec:Gate_Fidelity}
The operational space of the MAP gate spans the standard computational basis $\mathcal{C} = \{|000\rangle, |001\rangle, |100\rangle, |101\rangle\}$  alongside the higher-energy auxiliary states  $|\beta\rangle$  and  $|\alpha\rangle$. Particularly, a MAP gate can be performed by driving any of the eight transitions originating from the computational subspace. Furthermore, leakage dynamics are dominated by the nearest unintended transition, which varies based on the chosen intended drive and may even originate from the same computational state. Therefore, it is necessary to select a particular configuration for the leakage analysis, with the understanding that the derived mathematical framework applies equally to any intended transition and its corresponding nearest leakage channel. As our concrete example, we consider a gate that operates by accumulating a geometric phase via the $\vert{} 101\rangle \leftrightarrow \vert{} \beta\rangle$ transition. In this scenario, we analyze a prototypical leakage path wherein an unintended drive couples the computational state $\vert{}100\rangle$ to the auxiliary state $\vert{}\alpha\rangle$. 
To construct an accurate error model, we analyze a prototypical scenario wherein an unintended drive couples the computational state $|100\rangle$ to the auxiliary state $|\alpha\rangle$. 
Following the realization of the gate, the target state  $|101\rangle$  acquires the intended  $\pi$  phase, but a fractional population  $\eta$  leaks into  $|\alpha\rangle$. In the expanded basis  $\{|000\rangle, |001\rangle, |100\rangle, |101\rangle, |\alpha\rangle\}$, this dynamic is captured by the effective unitary matrix:
\begin{equation}
    U = \begin{pmatrix} 1 & 0 & 0 & 0 & 0 \\ 0 & 1 & 0 & 0 & 0 \\ 0 & 0 & \sqrt{1 - \eta} & 0 & -i\sqrt{\eta} \\ 0 & 0 & 0 & -1 & 0 \\ 0 & 0 & -i\sqrt{\eta} & 0 & \sqrt{1 - \eta} \end{pmatrix}.
\end{equation}
By modeling this leakage as a coherent process, the error channel can be fully characterized by a single Kraus operator, $K = U$. To compute the resulting fidelity, we compare this process to an ideal controlled-Z (CZ) operation,  $U_0$, augmented with an identity matrix applied to the leakage manifold: $U_0 = |000\rangle\langle000| + |001\rangle\langle001| + |100\rangle\langle100| - |101\rangle\langle101| + |\alpha\rangle\langle\alpha|$. Following the definition of average state fidelity in Ref.~\cite{Ding2023_MIT_FTF}, the corresponding leakage-induced contribution to the gate infidelity is estimated as 
\begin{equation}
   1- {\cal F} \simeq  \frac{1}{4}\eta + \frac{3}{80}\eta^2.
\end{equation}

\begin{acknowledgments}
This research was supported by the Guangdong Provincial Quantum Science Strategic Initiative (Grant No. GDZX2407001). We acknowledge insightful discussions with Xizheng Ma, Tenghui Wang, Jiahui Wang, Ran Gao, Liang Xiang, Wangwei Lan, and Hangxi Li.
\end{acknowledgments}

\bibliography{main.bib}

@article{Koch2007_transmon,
  title = {Charge-insensitive qubit design derived from the Cooper pair box},
  author = {Koch, Jens and Yu, Terri M. and Gambetta, Jay and Houck, A. A. and Schuster, D. I. and Majer, J. and Blais, Alexandre and Devoret, M. H. and Girvin, S. M. and Schoelkopf, R. J.},
  journal = {Phys. Rev. A},
  volume = {76},
  issue = {4},
  pages = {042319},
  numpages = {19},
  year = {2007},
  month = {Oct},
  publisher = {American Physical Society},
  doi = {10.1103/PhysRevA.76.042319},
  url = {https://link.aps.org/doi/10.1103/PhysRevA.76.042319}
}

@article{GoogleQAI_2019,
	title = {Quantum supremacy using a programmable superconducting processor},
	volume = {574},
	issn = {1476-4687},
	url = {https://doi.org/10.1038/s41586-019-1666-5},
	doi = {10.1038/s41586-019-1666-5},
	number = {7779},
	journal = {Nature},
	author = {{Google Quantum AI and Collaborators}},
	month = oct,
	year = {2019},
	pages = {505--510},
}

@misc{GoogleTunableCouplerPatent,
  author       = {Charles Neill and Anthony Edward Megrant},
  title        = {Tunable qubit coupler},
  howpublished = {\href{https://patents.google.com/patent/US11482656B2/en}{U.S. Patent US11482656 B2}},
  year         = {2022},
}

@article{Li2023_QCQ,
    doi = {10.1088/2058-9565/ace8b6},
    url = {https://doi.org/10.1088/2058-9565/ace8b6},
    year = {2023},
    month = {aug},
    publisher = {IOP Publishing},
    volume = {8},
    number = {4},
    pages = {045015},
    author = {Li, Fei-Yu and Jin, Li-Jing},
    title = {Quantum chip design optimization and automation in superconducting coupler architecture},
    journal = {Quantum Science and Technology}
}

@article{Krinner2022_IBMETH_d3,
	title = {Realizing repeated quantum error correction in a distance-three surface code},
	volume = {605},
	issn = {1476-4687},
	url = {https://doi.org/10.1038/s41586-022-04566-8},
	doi = {10.1038/s41586-022-04566-8},
	number = {7911},
	journal = {Nature},
	author = {Krinner, Sebastian and Lacroix, Nathan and Remm, Ants and Di Paolo, Agustin and Genois, Elie and Leroux, Catherine and Hellings, Christoph and Lazar, Stefania and Swiadek, Francois and Herrmann, Johannes and Norris, Graham J. and Andersen, Christian Kraglund and Müller, Markus and Blais, Alexandre and Eichler, Christopher and Wallraff, Andreas},
	month = may,
	year = {2022},
	pages = {669--674},
}

@article{GoogleQAI_2023,
	title = {Suppressing quantum errors by scaling a surface code logical qubit},
	volume = {614},
	issn = {1476-4687},
	url = {https://doi.org/10.1038/s41586-022-05434-1},
	doi = {10.1038/s41586-022-05434-1},
	pages = {676--681},
	number = {7949},
	journal = {Nature},
	author = {{Google Quantum AI and Collaborators}},
	date = {2023-02-01},
    year = 2023,
}

@article{GoogleQAI_2025,
	title = {Quantum error correction below the surface code threshold},
	volume = {638},
	issn = {1476-4687},
	url = {https://doi.org/10.1038/s41586-024-08449-y},
	doi = {10.1038/s41586-024-08449-y},
	number = {8052},
	journal = {Nature},
	author = {{Google Quantum AI and Collaborators}},
	month = feb,
	year = {2025},
	pages = {920--926},
}

@misc{Zuchongzhi2.1_2021,
  author       = {Qingling Zhu and others},
  title        = {Quantum Computational Advantage via 60-Qubit 24-Cycle Random Circuit Sampling},
  howpublished = {\href{https://arxiv.org/abs/2109.03494}{arXiv:2109.03494 [quant-ph]}},
  year         = {2021}
}

@article{Zuchongzhi3.2_2025,
  title = {Experimental Quantum Error Correction below the Surface Code Threshold via All-Microwave Leakage Suppression},
  author = {He, Tan and others},
  journal = {Phys. Rev. Lett.},
  volume = {135},
  issue = {26},
  pages = {260601},
  numpages = {7},
  year = {2025},
  month = {Dec},
  publisher = {American Physical Society},
  doi = {10.1103/rqkg-dw31},
  url = {https://link.aps.org/doi/10.1103/rqkg-dw31}
}

@article{Manucharyan2009_fluxonium,
    author = {Vladimir E. Manucharyan  and Jens Koch  and Leonid I. Glazman  and Michel H. Devoret},
    title = {Fluxonium: Single Cooper-Pair Circuit Free of Charge Offsets},
    journal = {Science},
    volume = {326},
    number = {5949},
    pages = {113-116},
    year = {2009},
    doi = {10.1126/science.1175552},
    url = {https://www.science.org/doi/abs/10.1126/science.1175552}
}

@article{Nguyen2019_fluxonium,
  title = {High-Coherence Fluxonium Qubit},
  author = {Nguyen, Long B. and Lin, Yen-Hsiang and Somoroff, Aaron and Mencia, Raymond and Grabon, Nicholas and Manucharyan, Vladimir E.},
  journal = {Phys. Rev. X},
  volume = {9},
  issue = {4},
  pages = {041041},
  numpages = {14},
  year = {2019},
  month = {Nov},
  publisher = {American Physical Society},
  doi = {10.1103/PhysRevX.9.041041},
  url = {https://link.aps.org/doi/10.1103/PhysRevX.9.041041}
}

@article{Somoroff2023,
  title = {Millisecond Coherence in a Superconducting Qubit},
  author = {Somoroff, Aaron and Ficheux, Quentin and Mencia, Raymond A. and Xiong, Haonan and Kuzmin, Roman and Manucharyan, Vladimir E.},
  journal = {Phys. Rev. Lett.},
  volume = {130},
  issue = {26},
  pages = {267001},
  numpages = {6},
  year = {2023},
  month = {Jun},
  publisher = {American Physical Society},
  doi = {10.1103/PhysRevLett.130.267001},
  url = {https://link.aps.org/doi/10.1103/PhysRevLett.130.267001}
}

@article{Fei2025,
  title = {High-coherence fluxonium qubits manufactured with a wafer-scale-uniformity process},
  author = {Wang, Fei and Lu, Kannan and Zhan, Huijuan and Ma, Lu and Wu, Feng and Sun, Hantao and Deng, Hao and Bai, Yang and Bao, Feng and Chang, Xu and Gao, Ran and Gao, Xun and Gong, Guicheng and Hu, Lijuan and Hu, Ruizi and Ji, Honghong and Ma, Xizheng and Mao, Liyong and Song, Zhijun and Tang, Chengchun and Wang, Hongcheng and Wang, Tenghui and Wang, Ziang and Xia, Tian and Xu, Hongxin and Zhan, Ze and Zhang, Gengyan and Zhou, Tao and Zhu, Mengyu and Zhu, Qingbin and Zhu, Shasha and Zhu, Xing and Shi, Yaoyun and Zhao, Hui-Hai and Deng, Chunqing},
  journal = {Phys. Rev. Appl.},
  volume = {23},
  issue = {4},
  pages = {044064},
  numpages = {14},
  year = {2025},
  month = {Apr},
  publisher = {American Physical Society},
  doi = {10.1103/PhysRevApplied.23.044064},
  url = {https://link.aps.org/doi/10.1103/PhysRevApplied.23.044064}
}

@article{Ding2023_MIT_FTF,
  title = {High-Fidelity, Frequency-Flexible Two-Qubit Fluxonium Gates with a Transmon Coupler},
  author = {Ding, Leon and Hays, Max and Sung, Youngkyu and Kannan, Bharath and An, Junyoung and Di Paolo, Agustin and Karamlou, Amir H. and Hazard, Thomas M. and Azar, Kate and Kim, David K. and Niedzielski, Bethany M. and Melville, Alexander and Schwartz, Mollie E. and Yoder, Jonilyn L. and Orlando, Terry P. and Gustavsson, Simon and Grover, Jeffrey A. and Serniak, Kyle and Oliver, William D.},
  journal = {Phys. Rev. X},
  volume = {13},
  issue = {3},
  pages = {031035},
  numpages = {24},
  year = {2023},
  month = {Sep},
  publisher = {American Physical Society},
  doi = {10.1103/PhysRevX.13.031035},
  url = {https://link.aps.org/doi/10.1103/PhysRevX.13.031035}
}

@misc{Zhan2026_22q_GHZ,
  author       = {Ze Zhan and Zishuo Li and Fei Wang and Wangwei Lan and Xianchuang Pan and Liang Xiang and Xu Dou and Ran Gao and Guicheng Gong and Yanbo Guo and Quan Guan and Lijuan Hu and Ruizhi Hu and Honghong Ji and Lijing Jin and Yongyue Jin and Chengyao Li and Kannan Lu and Lu Ma and Xizheng Ma and Hongcheng Wang and Jiahui Wang and Huijuan Zhan and Tao Zhou and Xing Zhu and Chunqing Deng and Tenghui Wang},
  title        = {Scalable Fluxonium Quantum Processors via Tunable-Coupler Architecture}, 
  howpublished = {\href{https://arxiv.org/abs/2604.13363}{arXiv:2604.13363 [quant-ph]}},
  year         = {2026}
}

@article{Rosenfeld2024,
  title = {High-Fidelity Two-Qubit Gates between Fluxonium Qubits with a Resonator Coupler},
  author = {Rosenfeld, Emma L. and Hann, Connor T. and Schuster, David I. and Matheny, Matthew H. and Clerk, Aashish A.},
  journal = {PRX Quantum},
  volume = {5},
  issue = {4},
  pages = {040317},
  numpages = {35},
  year = {2024},
  month = {Nov},
  publisher = {American Physical Society},
  doi = {10.1103/PRXQuantum.5.040317},
  url = {https://link.aps.org/doi/10.1103/PRXQuantum.5.040317}
}

@article{Heunisch2025,
  title = {Scalable fluxonium-transmon architecture for error-corrected quantum processors},
  author = {Heunisch, Lukas and Huang, Longxiang and Eckstein, Timo and Tasler, Stephan and Schirk, Johannes and Wallner, Florian and Feulner, Verena and Sarma, Bijita and Liegener, Klaus and Schneider, Christian M. F. and Filipp, Stefan and Hartmann, Michael J.},
  journal = {Phys. Rev. Res.},
  volume = {8},
  issue = {3},
  pages = {033245},
  numpages = {15},
  year = {2026},
  month = {Aug},
  publisher = {American Physical Society},
  doi = {10.1103/ts1j-nfg1},
  url = {https://link.aps.org/doi/10.1103/ts1j-nfg1}
}

@misc{Schirk2026, 
    author = {J. Schirk and N. Bruckmoser and S. M. Taubenberger and F. Wallner and N. J. Glaser and M. Zetzl and L. Huang and I. Tsitsilin and M. Werninghaus and L. Södergren and K. Liegener and C. M. F. Schneider and Stefan Filipp}, 
    title = {High-Fidelity Entangled States in a Connectivity-Four Fluxonium Quantum Processor}, 
    howpublished = {\href{https://arxiv.org/abs/2608.25503}{arXiv:2608.25503 [quant-ph]}}, year = {2026} 
}

@article{Peng2026,
  title = {Scalable native multiqubit gates via engineered noncomputational-state interactions in superconducting fluxonium qubits},
  author = {Zhao, Peng and Xu, Peng and Xue, Zheng-Yuan},
  journal = {Phys. Rev. A},
  volume = {113},
  issue = {2},
  pages = {022604},
  numpages = {12},
  year = {2026},
  month = {Feb},
  publisher = {American Physical Society},
  doi = {10.1103/xwny-vkft},
  url = {https://link.aps.org/doi/10.1103/xwny-vkft}
}

@misc{Chan2026_F_DTC,
  author       = {Guo Xuan Chan and Wangwei Lan and Tenghui Wang and Xizheng Ma and Chunqing Deng and Lijing Jin},
  title        = {System-Level Design of Scalable Fluxonium Quantum Processors with Double-Transmon Couplers}, 
  howpublished = {\href{https://arxiv.org/abs/2604.26373}{arXiv:2604.26373 [quant-ph]}},
  year         = {2026}
}

@article{Devoret1995_CQED,
  title={Quantum fluctuations in electrical circuits},
  author={Devoret, Michel H and others},
  journal={Les Houches, Session LXIII},
  volume={7},
  number={8},
  pages={133--135},
  year={1995},
  publisher={Elsevier New York}
}

@article{Blais2021_CQED,
  title = {Circuit quantum electrodynamics},
  author = {Blais, Alexandre and Grimsmo, Arne L. and Girvin, S. M. and Wallraff, Andreas},
  journal = {Rev. Mod. Phys.},
  volume = {93},
  issue = {2},
  pages = {025005},
  numpages = {72},
  year = {2021},
  month = {May},
  publisher = {American Physical Society},
  doi = {10.1103/RevModPhys.93.025005},
  url = {https://link.aps.org/doi/10.1103/RevModPhys.93.025005}
}

@article{Ding2021_CQED,
  title = {Free-mode removal and mode decoupling for simulating general superconducting quantum circuits},
  author = {Ding, Dawei and Ku, Hsiang-Sheng and Shi, Yaoyun and Zhao, Hui-Hai},
  journal = {Phys. Rev. B},
  volume = {103},
  issue = {17},
  pages = {174501},
  numpages = {17},
  year = {2021},
  month = {May},
  publisher = {American Physical Society},
  doi = {10.1103/PhysRevB.103.174501},
  url = {https://link.aps.org/doi/10.1103/PhysRevB.103.174501}
}

@article{Yan2018,
  title = {Tunable Coupling Scheme for Implementing High-Fidelity Two-Qubit Gates},
  author = {Yan, Fei and Krantz, Philip and Sung, Youngkyu and Kjaergaard, Morten and Campbell, Daniel L. and Orlando, Terry P. and Gustavsson, Simon and Oliver, William D.},
  journal = {Phys. Rev. Appl.},
  volume = {10},
  issue = {5},
  pages = {054062},
  numpages = {9},
  year = {2018},
  month = {Nov},
  publisher = {American Physical Society},
  doi = {10.1103/PhysRevApplied.10.054062},
  url = {https://link.aps.org/doi/10.1103/PhysRevApplied.10.054062}
}

@article{Goto2022_DTC,
  title = {Double-Transmon Coupler: Fast Two-Qubit Gate with No Residual Coupling for Highly Detuned Superconducting Qubits},
  author = {Goto, Hayato},
  journal = {Phys. Rev. Appl.},
  volume = {18},
  issue = {3},
  pages = {034038},
  numpages = {10},
  year = {2022},
  month = {Sep},
  publisher = {American Physical Society},
  doi = {10.1103/PhysRevApplied.18.034038},
  url ={https://link.aps.org/doi/10.1103/PhysRevApplied.18.034038}
}

@article{Li2024_DTC,
  title = {Realization of High-Fidelity CZ Gate Based on a Double-Transmon Coupler},
  author = {Li, Rui and Kubo, Kentaro and Ho, Yinghao and Yan, Zhiguang and Nakamura, Yasunobu and Goto, Hayato},
  journal = {Phys. Rev. X},
  volume = {14},
  issue = {4},
  pages = {041050},
  numpages = {30},
  year = {2024},
  month = {Nov},
  publisher = {American Physical Society},
  doi = {10.1103/PhysRevX.14.041050},
  url = {https://link.aps.org/doi/10.1103/PhysRevX.14.041050}
}

@misc{Tiwari2026_DTC,
  author       = {Tarush Tiwari and Sudhir K. Sahu and Guilhem Ribeill and Michael Senatore and Matthew D. LaHaye and Raymond W. Simmonds and Daniel L. Campbell and Archana Kamal and Leonardo Ranzani},
  title        = {High-fidelity iSWAP gate with Double Transmon Coupler}, 
  howpublished = {\href{https://arxiv.org/abs/2604.27080}{arXiv:2604.27080 [quant-ph]}},
  year         = {2026}
}

@misc{Peng2026_F_CapLd,
  author       = {Peng Zhao and Peng Xu and Zheng-Yuan Xue},
  title        = {Extensible Fluxonium Architecture Using Tunable Couplers with Low Shunt Capacitance}, 
  howpublished = {\href{https://arxiv.org/abs/2606.01647}{arXiv:2606.01647 [quant-ph]}},
  year         = {2026}
}

@article{Hertzberg2021_laser_annealing_JJ,
	title = {Laser-annealing {Josephson} junctions for yielding scaled-up superconducting quantum processors},
	volume = {7},
	issn = {2056-6387},
	url = {https://doi.org/10.1038/s41534-021-00464-5},
	doi = {10.1038/s41534-021-00464-5},
	number = {1},
	journal = {npj Quantum Information},
	author = {Hertzberg, Jared B. and Zhang, Eric J. and Rosenblatt, Sami and Magesan, Easwar and Smolin, John A. and Yau, Jeng-Bang and Adiga, Vivekananda P. and Sandberg, Martin and Brink, Markus and Chow, Jerry M. and Orcutt, Jason S.},
	month = aug,
	year = {2021},
	pages = {129},
}

@article{Najera-Santos2024_Grounded_Fluxonium,
  title = {High-Sensitivity ac-Charge Detection with a MHz-Frequency Fluxonium Qubit},
  author = {Najera-Santos, B.-L. and Rousseau, R. and Gerashchenko, K. and Patange, H. and Riva, A. and Villiers, M. and Briant, T. and Cohadon, P.-F. and Heidmann, A. and Palomo, J. and Rosticher, M. and le Sueur, H. and Sarlette, A. and Smith, W. C. and Leghtas, Z. and Flurin, E. and Jacqmin, T. and Del\'eglise, S.},
  journal = {Phys. Rev. X},
  volume = {14},
  issue = {1},
  pages = {011007},
  numpages = {18},
  year = {2024},
  month = {Jan},
  publisher = {American Physical Society},
  doi = {10.1103/PhysRevX.14.011007},
  url = {https://link.aps.org/doi/10.1103/PhysRevX.14.011007}
}

@article{Hida2025_Concentric_Fluxonium,
  title = {Flux-trapping fluxonium qubit},
  author = {Hida, Kotaro and Matsuura, Kohei and Watanabe, Shu and Nakamura, Yasunobu},
  journal = {Phys. Rev. Appl.},
  volume = {24},
  issue = {6},
  pages = {064069},
  numpages = {11},
  year = {2025},
  month = {Dec},
  publisher = {American Physical Society},
  doi = {10.1103/b99c-y22b},
  url = {https://link.aps.org/doi/10.1103/b99c-y22b}
}

@article{Pishchimova2023_JJ_Variation,
	title = {Improving {Josephson} junction reproducibility for superconducting quantum circuits: junction area fluctuation},
	volume = {13},
	issn = {2045-2322},
	url = {https://doi.org/10.1038/s41598-023-34051-9},
	doi = {10.1038/s41598-023-34051-9},
	number = {1},
	journal = {Scientific Reports},
	author = {Pishchimova, Anastasiya A. and Smirnov, Nikita S. and Ezenkova, Daria A. and Krivko, Elizaveta A. and Zikiy, Evgeniy V. and Moskalev, Dmitry O. and Ivanov, Anton I. and Korshakov, Nikita D. and Rodionov, Ilya A.},
	month = apr,
	year = {2023},
	pages = {6772},
}

@article{Kennedy2025_JJ_Variation,
  title = {Analysis of Josephson junction barrier variation: A combined electron microscopy, breakdown, and Monte Carlo approach},
  author = {Kennedy, Oscar W. and Crawford, Kevin G. and Shahbazi, Kowsar and Shelly, Connor D.},
  journal = {Phys. Rev. Mater.},
  volume = {9},
  issue = {8},
  pages = {084803},
  numpages = {10},
  year = {2025},
  month = {Aug},
  publisher = {American Physical Society},
  doi = {10.1103/5tb4-pslg},
  url = {https://link.aps.org/doi/10.1103/5tb4-pslg}
}

@article{Shapovalov2020_lowCJJ,
  title={Low-capacitance Josephson junctions},
  author={Shapovalov, AP and Febvre, Pascal and Yilmaz, U and Shnyrkov, VI and Belogolovskii, MO and Kordyuk, OA},
  journal={Progress in physics of metals},
  volume={21},
  number={1},
  pages={3--25},
  year={2020},
  url={https://cnrs.hal.science/hal-04873659/file/UspFizMet_2020.pdf}
}

@article{Watanabe2003_smallCJJ,
  title = {Quantum effects in small-capacitance single Josephson junctions},
  author = {Watanabe, Michio and Haviland, David B.},
  journal = {Phys. Rev. B},
  volume = {67},
  issue = {9},
  pages = {094505},
  numpages = {11},
  year = {2003},
  month = {Mar},
  publisher = {American Physical Society},
  doi = {10.1103/PhysRevB.67.094505},
  url = {https://link.aps.org/doi/10.1103/PhysRevB.67.094505}
}

@article{chen2021_repetition,
	title = {Exponential suppression of bit or phase errors with cyclic error correction},
	volume = {595},
	issn = {1476-4687},
	url = {https://doi.org/10.1038/s41586-021-03588-y},
	doi = {10.1038/s41586-021-03588-y},
	number = {7867},
	journal = {Nature},
	author = {Chen, Zijun and {Google Quantum AI} and others},
	month = jul,
	year = {2021},
	pages = {383--387},
}

@article{Wang2026_qLDPC,
	title = {Demonstration of low-overhead quantum error correction codes},
	volume = {22},
	issn = {1745-2481},
	url = {https://doi.org/10.1038/s41567-025-03157-4},
	doi = {10.1038/s41567-025-03157-4},
	number = {2},
	journal = {Nature Physics},
	author = {Wang, Ke and others},
	month = feb,
	year = {2026},
	pages = {308--314},
}

@article{Zhuang_2026_Fluxonium_Loss,
	title = {Non-{Markovian} relaxation spectroscopy of fluxonium qubits},
	volume = {17},
	issn = {2041-1723},
	url = {https://doi.org/10.1038/s41467-026-69910-2},
	doi = {10.1038/s41467-026-69910-2},
	number = {1},
	journal = {Nature Communications},
	author = {Zhuang, Ze-Tong and Rosenstock, Dario and Liu, Bao-Jie and Somoroff, Aaron and Manucharyan, Vladimir E. and Wang, Chen},
	month = feb,
	year = {2026},
	pages = {3209},
}

@article{Chamberland2020_IBMColorCode,
doi = {10.1088/1367-2630/ab68fd},
url = {https://doi.org/10.1088/1367-2630/ab68fd},
year = {2020},
month = {feb},
publisher = {IOP Publishing},
volume = {22},
number = {2},
pages = {023019},
author = {Chamberland, Christopher and Kubica, Aleksander and Yoder, Theodore J and Zhu, Guanyu},
title = {Triangular color codes on trivalent graphs with flag qubits},
journal = {New Journal of Physics}
}

@article{Breuckmann2021_QLDPC,
  title = {Quantum Low-Density Parity-Check Codes},
  author = {Breuckmann, Nikolas P. and Eberhardt, Jens Niklas},
  journal = {PRX Quantum},
  volume = {2},
  issue = {4},
  pages = {040101},
  numpages = {19},
  year = {2021},
  month = {Oct},
  publisher = {American Physical Society},
  doi = {10.1103/PRXQuantum.2.040101},
  url = {https://link.aps.org/doi/10.1103/PRXQuantum.2.040101}
}

\end{document}